\documentclass[prx,reprint, longbibliography,superscriptaddress]{revtex4-2}
\usepackage{braket}
\usepackage{graphicx}
\usepackage{amsmath}
\usepackage{amssymb}
\usepackage{hyperref}
\usepackage[all]{hypcap}
\usepackage[dvipsnames]{xcolor}
\usepackage{comment}
\usepackage{bbm}

\definecolor{dblue}{rgb}{0.01, 0.33, 1.0}
\hypersetup{colorlinks=true,allcolors=dblue}
\usepackage{placeins}
\usepackage{times}
\usepackage{bbold}
\DeclareMathOperator{\Tr}{Tr}

\begin{document}
\author{Xin H. H. Zhang}
\email[Corresponding author: ]{physicsxinzhang@gmail.com}
\affiliation{Technical University of Munich, TUM School of Natural Sciences, Physics Department, 85748 Garching, Germany}
\affiliation{Walther-Mei{\ss}ner-Institut, Bayerische Akademie der Wissenschaften, 85748 Garching, Germany}
\affiliation{Munich Center for Quantum Science and Technology (MCQST), 80799 Munich, Germany}

\author{Daniel Malz}
\affiliation{Department of Mathematical Sciences, University of Copenhagen, 2100 Copenhagen, Denmark}

\author{Peter Rabl}
\affiliation{Technical University of Munich, TUM School of Natural Sciences, Physics Department, 85748 Garching, Germany}
\affiliation{Walther-Mei{\ss}ner-Institut, Bayerische Akademie der Wissenschaften, 85748 Garching, Germany}
\affiliation{Munich Center for Quantum Science and Technology (MCQST), 80799 Munich, Germany}

\date{July 28, 2026}

\title{Robust Superradiance and Spontaneous Spin Ordering in Disordered Waveguide QED}

\title{Robust Superradiance and Spontaneous Spin Ordering in Disordered \\ Waveguide Quantum Electrodynamics}

\begin{abstract}

We study the collective emission of a disordered array of $N$ excited two-level atoms into a one-dimensional photonic waveguide. In the perfectly ordered case, where atoms are spaced by exact integer multiples of the wavelength, the system exhibits the characteristic superradiant burst with a peak emission rate scaling as $N^2$. Using large-scale semiclassical simulations, we find that this key signature of superradiance remains asymptotically robust under strong spatial and spectral disorder, but also exhibits subtle finite-size scaling toward this limit. To explain our observations, we provide an analytical variational estimate for the maximal decay rate, which tightly bounds the numerical results and reveals how disorder shapes the collective decay. Specifically, we find that even in the presence of strong disorder, the spins tend to self-organize spontaneously according to their locations, which overall optimizes constructive interference effects and explains the emergence of mirror-asymmetric correlations in superradiant decay. These findings resolve important open questions regarding the existence and nature of superradiance in strongly disordered arrays and offer valuable insights for understanding collective quantum optical phenomena in realistic systems.

\end{abstract}

\maketitle


\section{Introduction}

Collective spontaneous emission, or superradiance, is a hallmark of cooperative quantum behavior, first predicted by Dicke in 1954~\cite{DickePhysRev1954}. It was shown that the collective emission of a set of $N$ two-level atoms (TLAs) or other emitters can be drastically accelerated by the build-up of correlations between individual dipoles. In Dicke’s idealized model, with symmetric all-to-all interactions, this effect leads to a sharp superradiant photon burst at a time $\sim \ln(N)/(N\gamma)$, with a peak emission rate proportional to $\gamma N^2$, where $\gamma$ is the decay rate of an individual emitter. Since then, this phenomenon has sparked extensive theoretical~\cite{RehlerPRA1971,BonifacioPRA1971,BonifacioPRA1971b,DegiorgioOC1971,DegiorgioPRA1971,HaakePRA1972,NarducciPRA1974,Agarwal1974,BonifacioPRA1975,MacGillivrayPRA1976,HaakePRL1979,HaakePRA1979,SchuurmansOC1980,MattarPRL1981,DrummondPRA1982,GrossPhysRep1982} and experimental~\cite{SkribanowitzPRL1973,GrossPRL1976,VrehenPRL1977,RaimondPRL1982} investigations, and remains a cornerstone of quantum optics.

Recently, there has been a revived interest in superradiance~\cite{GobanPRL2015,ZhuNJP2015,AsenjoGarciaPRX2017,WangPRL2020,BrehmNatMat2021,PineiroPRX2022,MassonNatComm2022,MalzPRA2022,ReitzPRXQ2022,OriolPRR2023,MokPRL2023,FerioliNatPhys2023,MinkSciPostPhys2023,CardenasLopezPRL2023,TiranovScience2023,LiedlPRX2024,FasserOQ24,TebbenjohannsPRA2024,BachArXiv2024,WindtPRL2025,MokPRR2025,ZhangPRL2025,LeeArxiv2025,VovkArxiv2025,MokArxiv2024,HolzingerArxiv2025,BasslerArxiv2025,RosarioPRL2025}, mainly stimulated by the development of new experimental platforms, in which superradiant effects can be explored at an unprecedented level of control, and potentially be exploited in applications such as superradiant lasers~\cite{MeiserPRL2009,BohnetNature2012,NorciaSciAdv2016} or quantum metrology~\cite{WangPRL2014,PaulischPRA2019,PerarnauLlobetQST2020,Arya2024,BelliardoArxiv2026}.
Of particular relevance in this context are waveguide QED systems~\cite{RoyRMP2017,ChangRMP2018,SheremetRMP2023}, where the emitted photons are confined to a one-dimensional (1D) channel and atom-atom correlations can be established efficiently and over long distances~\cite{GobanPRL2015,AsenjoGarciaPRX2017,WangPRL2020,CardenasLopezPRL2023,TiranovScience2023,LiedlPRX2024,FasserOQ24, TebbenjohannsPRA2024,BachArXiv2024,WindtPRL2025,MokPRR2025,ZhangPRL2025,LeeArxiv2025,VovkArxiv2025}.
Specifically, when emitters are placed in a lattice with a spacing that matches an exact multiple of the resonant wavelength, all the emitted photons can interfere fully constructively, thereby recovering the features of the original Dicke model in an extended setting.  
However, in real systems, the positions of the atoms as well as their frequencies fluctuate, giving rise to competing interference patterns once the disorder becomes sufficiently strong. This raises an important fundamental question: Does Dicke superradiance survive in disordered quantum systems, and how does disorder affect its characteristic features?

\begin{figure}[t]
	\center
\includegraphics[width=\columnwidth]{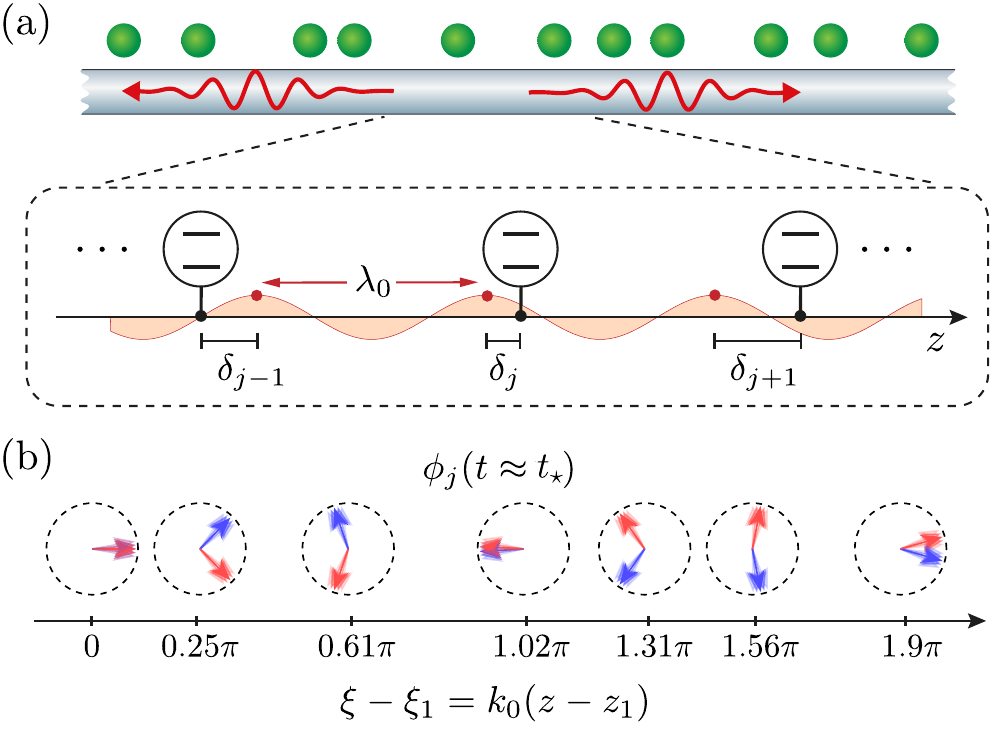}
	\caption{(a) Sketch of a disordered waveguide QED system, where $N$ excited TLAs are located at random positions $z_j$ along a 1D photonic channel and undergo collective spontaneous decay. As shown in the inset, the displacements $\delta_j$ from exact multiples of the resonant wavelength $\lambda_0$ are chosen randomly from the interval $\left[-\Theta/(2k_0),\Theta/(2k_0)\right)$ to interpolate between a regular array ($\Theta=0)$ and a uniformly distributed gas of atoms ($\Theta\gg1$). (b) Illustration of the two ordering patterns (red and blue), which are generated spontaneously between the atomic dipoles during the superradiant decay. The dipoles self-align according to their distances, with $\phi_{j} \approx \phi_{1}  + k_{0} (z_{j} - z_{1})$ (blue arrows) or $\phi_{j} \approx \phi_{1}  - k_{0} (z_{j} - z_{1})$ (red arrows), which even for strong disorder yields constructive interference and enhanced collective emission either to the left or to the right.}
	\label{FigSchematic}
\end{figure}

While it is generally expected that collective radiation effects are relatively robust against weak disorder, the fate of superradiant emission under strong disorder remains unclear. This concerns, in particular, the scaling behavior of the superradiant burst in 1D, where superradiance could be rather robust due to long-range interactions on the one hand, while on the other hand, the influence of competing interference patterns or disorder-induced localization is most pronounced. In the studies of pencil-shaped atomic gases~\cite{GrossPhysRep1982,SkribanowitzPRL1973,HaakePRL1979,HaakePRA1979,SchuurmansOC1980,MattarPRL1981,DrummondPRA1982,MacGillivrayPRA1976}, which can be approximated as a 1D model, superradiance has been shown with the atoms treated as a continuous medium and without quantum fluctuations in the decaying dynamics. It remains unclear if superradiance can exist in a fully quantum model and beyond the continuum limit. Recently, the existence of superradiance in disordered systems has been inferred from initial photon correlations~\cite{MassonNatComm2022,CardenasLopezPRL2023}, upper bounds on the maximal decay rate~\cite{MokArxiv2024}, and heuristic scaling arguments~\cite{HolzingerArxiv2025}, but the validity of these arguments and the actual scaling of the peak emission rate remain poorly understood. In particular, it is unclear whether the $N^2$ scaling survives in the presence of strong disorder, and if so, how the atoms can sustain collective behavior despite competing phase relations that disrupt spatial correlations and a global phase alignment.

Without permutation symmetry, exact numerical simulations of superradiance quickly become intractable. To overcome this limitation, here we use scalable semi-classical methods to perform a systematic study of superradiant emission in a disordered array of atoms coupled to a 1D waveguide [see Fig.~\ref{FigSchematic}(a)].
To obtain both quantitatively accurate results as well as an intuitive physical understanding, we combine large-scale numerical simulations based on the discrete truncated Wigner approximation (DTWA)~\cite{SchachenmayerPRX2015,HuberPRA2022,MinkSciPostPhys2023} and on a quantum state diffusion (QSD) approach~\cite{GisinJPA1992, Carmichael1993,ZhangPRL2025} with an approximate analytical analysis of the decay rates. Both our numerical and analytic approaches are supported by recent findings on separable descriptions of superradiance~\cite{ZhangPRL2025,RosarioPRL2025,BasslerArxiv2025}, and are justified independently through numerical benchmarks against exact results.

By investigating the maximal emission rate and the peak emission time for an increasing number of atoms, we demonstrate that even for strong disorder, the Dicke scaling is asymptotically recovered. There are, however, subtle differences in the numerical prefactors and in how this limit is approached. 
To explain these observations, we introduce a product-state ansatz for the configuration of atomic dipoles at the time of the emission peak, which provides a tight upper bound for the maximal collective decay rate. 
Remarkably, this bound is not only close to the numerically exact values, but also reveals a simple and intuitive mechanism for robustness: the atomic spins organize spontaneously in random directions, but with their relative orientations being determined by their relative locations along the waveguide [Fig.~\ref{FigSchematic}(b)]. This emergent spin ordering, without any external control, enables constructive interference in the decay channels for any atomic configuration, which enables the persistence of the $N^2$ scaling regardless of the disorder strength.

The paper is organized as follows. In Sec.~\ref{secSetup}, we introduce the model: a disordered chain of atoms coupled to a 1D photonic waveguide. Then, Sec.~\ref{secNumerics} outlines the numerical methods used in our study, including benchmarks against exact simulations and a brief comparison with various other methods. In Sec.~\ref{secNumericalResults}, we present systematic large-scale simulations demonstrating that the characteristic $N^2$ superradiant scaling remains robust under strong spatial disorder. To gain analytical insight, Sec.~\ref{SecUpperBoundR} derives an upper bound on the collective decay rate, revealing a form of spontaneous spin ordering that underlies this robustness. In Sec.~\ref{secGeneral}, we also show that in the large-$N$ limit, superradiance becomes robust against spectral disorder and that it prevails in non-Markovian settings. Finally, we conclude with a summary of our main findings in Sec.~\ref{secConclusion}.


\section{Model}\label{secSetup}

We consider a set of $N$ TLAs, which are located at fixed positions $z_j$ along a 1D photonic waveguide [see Fig.~\ref{FigSchematic}(a)]. The TLAs have a ground state $|g\rangle$ and an excited state $|e\rangle$, which are separated by a transition frequency $\omega_{0}$. Under the assumption that over a sufficiently broad frequency range around $\omega_0$ the waveguide exhibits an approximately linear dispersion relation with group velocity $v_{\rm g}$, we can trace out the waveguide modes and describe the remaining dynamics of the atoms by a master equation of the form (see, for example,~\cite{DzsotjanPRB2010,LalumierePRA2013})
\begin{equation} \label{ME0}
 \dot  \rho = -i [H, \rho] +  \frac{\gamma}{2}  \mathcal{D}[J_{\text{R}}] \rho + \frac{\gamma}{2}  \mathcal{D}[J_{\text{L}}] \rho.
\end{equation}
Here $H$ accounts for coherent, photon-mediated interactions between the atoms, 
\begin{equation}\label{Hamiltonian}
    H = \frac{\gamma}{2} \sum_{i,j} \sin(k_{0} |z_{i} - z_{j}|) \sigma_{i}^{+} \sigma_{j}^{-},
\end{equation}
where $\sigma_{j}^{+}$ ($\sigma_{j}^{-}$) denotes the raising (lowering) operator for the $j$-th atom, $k_{0} = \omega_{0}/v_{\rm g}$ is the resonant wavevector and $\gamma$ the single-atom decay rate.  Incoherent losses of excitations are described by the last two terms in Eq.~\eqref{ME0}, where we introduced the dissipator $\mathcal{D}[J] \bullet = J \bullet J^{\dagger} - \frac{1}{2} \{ \bullet, J^{\dagger} J \}$ and the two collective jump operators   
\begin{equation} \label{JumpOp}
    J_{\text{R/L}} =  \sum_{j=1}^N e^{\mp i k_{0} z_{j}} \sigma_{j}^{-},
\end{equation}
describing emission into right- and left-propagating modes, respectively \footnote{Note that for a chiral waveguide, the spatial dependence can be removed with a phase transformation for the spin operators.}.
We are primarily interested in the superradiant dynamics of the system starting from the fully excited state $\rho(0)=\ket{e\dots e}\bra{e\dots e}$, which we will use as an initial state for all the results presented below. 
In our analysis, we neglect local dephasing and additional loss channels into other nonguided modes, since any additional dissipation with a rate comparable to $\gamma$ has little effect on superradiance, which occurs on a time scale $\sim \ln(N)/(\gamma N)$.

From Eq.~\eqref{Hamiltonian} and Eq.~\eqref{JumpOp}, we see that as a result of interference effects, both the coherent interaction and the collective dissipation terms depend strongly on the positions of the emitters. Only in a perfectly ordered array, for example with $z_j= j \times \lambda_0$, where $\lambda_0=2\pi/k_0$ is the wavelength of the emitted photons, all the phase factors drop out and we recover the usual master equation for Dicke superradiance with $H=0$ and homogeneous jump operators $J_{\text{R}}=J_{\text{L}}= \sum_{j} \sigma_{j}^{-}$. In the following, we are interested in disordered configurations where the atoms are offset from the ideal periodic array by an amount $\delta_j$, i.e., 
\begin{equation}
z_{j} = j \times \lambda_0 + \delta_{j}.
\end{equation} 
In our study, the offset phases $\xi_{j} = k_{0} \delta_{j}$ are randomly chosen from a uniform distribution in the interval $ [-\Theta/2, \Theta/2)$, such that $\Theta$ quantifies the degree of disorder. Note that for this configuration, neighboring atoms can overlap when $\Theta > 2\pi$ and in the limit $\Theta \rightarrow \infty$ we recover a fully homogeneous density distribution spatially. The dynamics is determined by the distribution of the relative propagation phases, $\Delta \xi_{ij}=(\xi_i-\xi_j)\mod 2\pi$, which obeys a wrapped triangular distribution (see the Supplemental Material \cite{SupMat}) and approaches a uniform distribution for any $\Theta = 2 n\pi$ with $n \in \mathbb{Z}_+$ and in the limit $\Theta\rightarrow \infty$.
Additional simulations for an ensemble with Gaussian-distributed $\xi_j$ are performed for comparison, ensuring that the general conclusions do not depend on any specifics of the disorder. To connect the results to realistic platforms, in \cite{SupMat}, we also discuss the expected range of the disorder strength $\Theta$ for a representative set of experimental systems with atomic and solid-state emitters.


Due to the bidirectional nature, by diagonalizing the decay rate matrix (see Appendix \ref{SecSingleExcitationUpperBound}), the master equation in Eq.~\eqref{ME0} can be equivalently written as 
\begin{equation}\label{TwoChannelLindbladian}
    \frac{d}{dt} \rho = -i [H, \rho] + \Gamma_{+} \mathcal{D}[J_+] \rho + \Gamma_{-} \mathcal{D}[J_{-}] \rho,
\end{equation}
where $J_{\pm}$ are collective jump operators and $ \Gamma_{\pm} = \gamma N ( 1 \pm |c|)/2$ are the corresponding decay rates.
Here, we have introduced the complex parameter 
\begin{equation}\label{cDef}
c =  \frac{1}{N}\sum_{j} e^{-i2\xi_{j}},
\end{equation}
which characterizes the overlap between the original left- and right-going decay modes (see Appendix \ref{SecSingleExcitationUpperBound}). It assumes a value of $c=1$ in the absence of disorder, i.e., $\xi_{j}=0$, and a value of $c\simeq 0$ when the atoms are homogeneously distributed. The short-time dynamics from the fully excited state $\rho(0)$ can be described in a linearized picture, in which the two jump operators $J_\pm$ act independently.
At a later stage, the linearized picture is no longer appropriate, since the states generated by multiple applications of $J_{\pm}$ are not orthogonal.

\section{Numerical methods}\label{secNumerics}

As mentioned above, in the absence of disorder, the Hamiltonian vanishes, $H=0$, and the jump operators preserve the total angular momentum. For the case of an initially fully excited ensemble of atoms, the decay dynamics thus reduces to a rate equation in a $(N+1)$-dimensional subspace, which can be solved numerically for very large $N$. For the general case with $\delta_i\neq 0$, this is no longer the case, and an exact integration of the full master equation is only possible for very small $N$. To overcome this problem, here we implement two approximate semiclassical methods based on the DTWA~\cite{SchachenmayerPRX2015,HuberPRA2022} and on QSD simulations~\cite{GisinJPA1992,Carmichael1993}.

\subsection{Discrete Truncated Wigner Approximation}\label{subsecDTWA}

\begin{figure}[t]
	\center  
\includegraphics[width=\columnwidth]{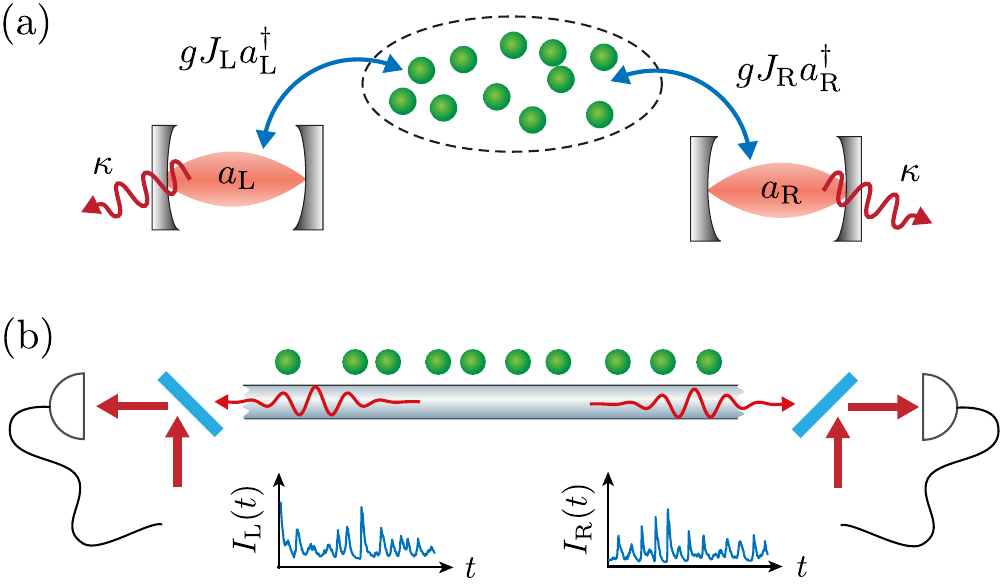}    
	\caption{Numerical methods. (a) Sketch of the extended system described by Eq.~\eqref{extendedME}, where the collective decay into left- and right-going waveguide modes is realized effectively via the coupling of the atoms to two lossy cavity modes with decay rate $\kappa$. This extended model underlies the DTWA method discussed in Sec.~\ref{subsecDTWA} and is used for the study of non-Markovian effects in Sec.~\ref{secNonMarkov}.  In the Markovian limit $\kappa/(\gamma N)  \to \infty$, the model recovers the original Lindblad master equation in Eq.~\eqref{ME0}. (b) Sketch of a heterodyne measurement setup that corresponds to the QSD unraveling of the master equation, as discussed in Sec.~\ref{secQSD}. Trajectories obtained in our QSDMF simulations represent (factorized) states of the atoms that are conditioned on a given history of the measured heterodyne currents $I_{\rm R}(t)$ and $I_{\rm L}(t)$.  }
	\label{FigCavityDecay}
\end{figure}

In our first approach, we make use of the fact that in a 1D system, photons can only be emitted to the left or to the right. This implies that dissipation in a waveguide can equivalently be modeled in terms of a coherent coupling of the atoms to two independent bosonic modes, which are themselves heavily damped [see Fig.~\ref{FigCavityDecay}(a)]. Therefore, in the following, we remodel the waveguide setup by an extended master equation of the form 
\begin{equation}\label{extendedME}
\dot \rho= -i [ H+ H_{\text{int}}, \rho ]+ \kappa \mathcal{D}[a_{\text{R}}]\rho + \kappa \mathcal{D}[a_{\text{L}}]\rho,
\end{equation} 
where $a_{\text{R/L}}$ ($a_{\text{R/L}}^\dagger$) are bosonic annihilation (creation) operators and 
\begin{equation}
    H_{\rm int}= i\frac{g}{\sqrt{2}} \left( J_{\text{R}} a^\dagger_{\text{R}} +  J_{\text{L}} a_L^\dagger - {\rm H.c.} \right)
\end{equation}
is the atom-photon interaction with a coupling strength $g$. From this extended model, the Markovian master equation for the atoms in Eq.~\eqref{ME0} is recovered by taking the limit $\kappa/(\gamma N) \rightarrow \infty$, while keeping $\gamma=4 g^2/\kappa$ fixed. This mapping can be straightforwardly generalized to study generic dissipative channels. 

For this enlarged system, we follow Ref.~\cite{HuberPRA2022} and perform a truncated Wigner approximation for both the atomic and bosonic degrees of freedom. In this approximation, each TLA is modeled in terms of a classical spin vector $\vec s_j=(s_j^x,s_j^y,s_j^z)$ and the bosonic modes in terms of complex amplitudes $\alpha_{\text{R}}$ and $\alpha_{\text{L}}$. These classical variables evolve according to the mean-field (MF) equations of motion (EOM) derived from Eq.~\eqref{extendedME}. For the current setup, these equations read 
\begin{equation}
\begin{split}
    \label{SpinEOM}
        \frac{d}{dt} s_{j}^{-} &= i \frac{\gamma}{2}\sum_{l\neq j} \sin(|\xi_j-\xi_l|) s_j^z s_l^- \\
        &+ \frac{g}{\sqrt{2}} s_{j}^{z} \left( \alpha_{\text{R}} e^{i\xi_j}  +  \alpha_{\text{L}} e^{-i\xi_j} \right)  , \\
        \frac{d}{dt} s_{j}^{z} &= - \gamma  \sum_{l \neq j} \sin(|\xi_{j} - \xi_{l}|) ( i s_{j}^{+} s_{l}^{-} - i s_{j}^{-} s_{l}^{+}  )   \\
        & - \sqrt{2} g s_j^+ \left(  \alpha_{\text{R}} e^{i\xi_j}  + \alpha_{\text{L}} e^{-i\xi_j}  \right)  - {\text{c.c.}},  
\end{split}
\end{equation}
for the spin variables with $s_j^-=(s_j^+)^*=(s_j^x - i s_j^y)/2$, and 
\begin{equation}
    \label{AlphaSDE}
\begin{split}
        d\alpha_{\text{R}} &= - \frac{\kappa}{2} \alpha_{\text{R}} dt + \frac{g}{\sqrt{2}}  \tilde{J}_{\text{R}} dt + \sqrt{\frac{\kappa}{2}} dW_{\text{R}}, \\ 
        d\alpha_{\text{L}}  &= - \frac{\kappa}{2} \alpha_{\text{L}}  dt +  \frac{g}{\sqrt{2}}  \tilde{J}_{\text{L}} dt + \sqrt{\frac{\kappa}{2}} dW_{\text{L}} ,
\end{split}
\end{equation}
for the amplitudes. Here, $\tilde{J}_{\text{R/L}} = \sum_{j=1}^N e^{\mp i k_{0} z_{j}} s_{j}^{-} $, and  $dW_{\text{R/L}}$ are two independent complex Wiener increments.

While Eq.~\eqref{SpinEOM} and Eq.~\eqref{AlphaSDE} describe a purely classical evolution, the effect of quantum fluctuations is approximately taken into account by sampling the initial values of the spin variables from a probability distribution. Specifically, for the spins, we have a discrete initial distribution with values 
\begin{equation}
\vec s_j= (\pm1,\pm1, 1)
\end{equation}
occurring with equal probability of $1/4$. This distribution corresponds to the case of an atom in state $\ket{e}$, and reproduces the correct quantum mechanical expectation values $\braket{\sigma^x} =\braket{ \sigma^y}=0$, $\braket{ \sigma^z}=1$ and $\braket{(\sigma^{\mu})^2}=1$ when averaging over all configurations. For the bosonic modes, the initial values for $\alpha_R$ and $\alpha_L$ are sampled from the Gaussian initial distribution 
\begin{equation}
W_0(\alpha)= \frac{2}{\pi} e^{-2|\alpha|^2}, 
\end{equation}
which corresponds to the Wigner function of the vacuum state. Since the bosonic modes decay, an appropriate amount of noise must be added during the evolution to preserve the minimum quantum uncertainty. This is taken into account by the stochastic terms in Eq.~\eqref{AlphaSDE}.

The set of stochastic equations in Eq.~\eqref{SpinEOM} and Eq.~\eqref{AlphaSDE} can be numerically integrated in a straightforward manner, and the relevant expectation values of interest can be obtained by averaging over a sufficiently large number of trajectories $\mathcal{N}$. For example, 
\begin{eqnarray}
\langle \sigma_i^{\mu} \sigma_j^{\nu} \rangle (t)  &\simeq& \frac{1}{\mathcal{N}} \sum_n s_i^{\mu}(t) s_j^{\nu}(t), \\
\langle (a_{\text{R}}^\dag)^k a_{\text{R}}^m \rangle_{\rm sym}(t) &\simeq& \frac{1}{\mathcal{N}} \sum_n [\alpha_{\text{R}}^*(t)]^k\alpha_{\text{R}}^m(t),
\end{eqnarray}
where $\langle \bullet \rangle_{\rm sym}$ denotes the symmetrically ordered expectation values. To recover the original master equation, we find that $\kappa \gtrsim 50 \gamma N$ is sufficient to eliminate non-Markovian effects in our simulations. Alternatively, as described in Appendix~\ref{DTWAappendix}, the cavity amplitudes can be integrated out analytically to derive a set of stochastic equations of motion of the spin variables only, similar to the derivation performed in Ref.~\cite{HosseinabadiPRXQ2025} using a path integral approach. However, a key feature of this method is that the assumption of Markovianity can be lifted by simply choosing smaller values of $\kappa$. In this way, collective emission effects can also be simulated in non-Markovian scenarios without additional computational cost (see Sec.~\ref{secNonMarkov}).

\subsection{Quantum State Diffusion}\label{secQSD}
As a closely related, but in several aspects complementary method, we also simulate the decay dynamics using a QSD unraveling of the original master equation in Eq.~\eqref{ME0}, combined with a product-state ansatz for each individual trajectory. This mean-field QSD (``QSDMF") approach is motivated by the recent study in Ref.~\cite{ZhangPRL2025}, which argues that this method becomes essentially exact for simulating superradiance in the large-$N$ limit. Compared to DTWA, QSDMF offers access to physical quantum trajectories that reveal measurement-induced phenomena and provide additional insights at the trajectory level.  

As depicted in Fig.~\ref{FigCavityDecay}(b), QSD describes the dynamics of the monitored waveguide system under heterodyne measurement of the emitted field amplitudes. In this formalism, the master equation in Eq.~\eqref{ME0} is unraveled into an ensemble of pure trajectories obeying the QSD equation~\cite{GisinJPA1992, Carmichael1993}
\begin{align}\label{QSDgeneral}
    d\ket{\psi} = & - \frac{\gamma}{4} \sum_{k} \left( J_{k}^{\dagger} J_{k} + \braket{J_{k}}^{*} \braket{J_{k}} - 2 \braket{J_{k}}^{*} J_{k} \right) \ket{\psi} dt \nonumber \\ &+ \sum_{k} \sqrt{\frac{\gamma}{2}}(J_{k} - \braket{J_{k}}) \ket{\psi} dW_{k} -i H \ket{\psi} dt,
\end{align}
with $k=\text{R, L}$ and $dW_k$ two complex-valued Wiener increments. Each trajectory represents the state of the system conditioned on the heterodyne currents 
\begin{equation}
dI_{k}(t) = \frac{\gamma}{2} \langle J_k\rangle dt + \sqrt{\frac{\gamma}{2}} dW_k  
\end{equation}
measured at the left and right outputs. Unconditioned expectation values of observables are obtained from averages over the trajectories.

In general, the wavefunction in Eq.~\eqref{QSDgeneral} still scales exponentially with the atom number, and to make this method tractable, we approximate each trajectory by a product state, $|\psi\rangle\approx\otimes_{j=1}^N |\psi_j\rangle$. This simplification allows us to describe each trajectory again in terms of classical spin variables $\vec{\mathfrak{s}}_j(t)$, with components $\mathfrak{s}^\eta_j=\langle \psi_j|\sigma_j^\eta|\psi_j\rangle$. The corresponding stochastic equations for these spin vectors for the current setting are given in Ref.~\cite{ZhangPRL2025}. Note that QSDMF can be improved by refining the treatment of individual trajectories. For example, including second-order moments~\cite{VerstraelenPRXQ2023,LiNatComm2025} in the EOM of trajectories can encode some quantum correlations, at the cost of a less favorable scaling $\mathcal{O}(N^2)$. Importantly, by treating trajectories using matrix product states (MPS) \cite{ZhangPRL2025}, it can be systematically improved by increasing the bond dimension $d$, with complexity $\mathcal{O}(N d^3)$. This method is thus particularly efficient if entanglement in each trajectory is constant or logarithmic in $N$, which requires $d\sim \mathcal{O}(1)$ or $d\sim \mathcal{O}(N)$, respectively, for giving numerically exact results.


\subsection{Benchmarking}
\label{secValidity}

\begin{figure}[t]
	\center
\includegraphics[width=\columnwidth]{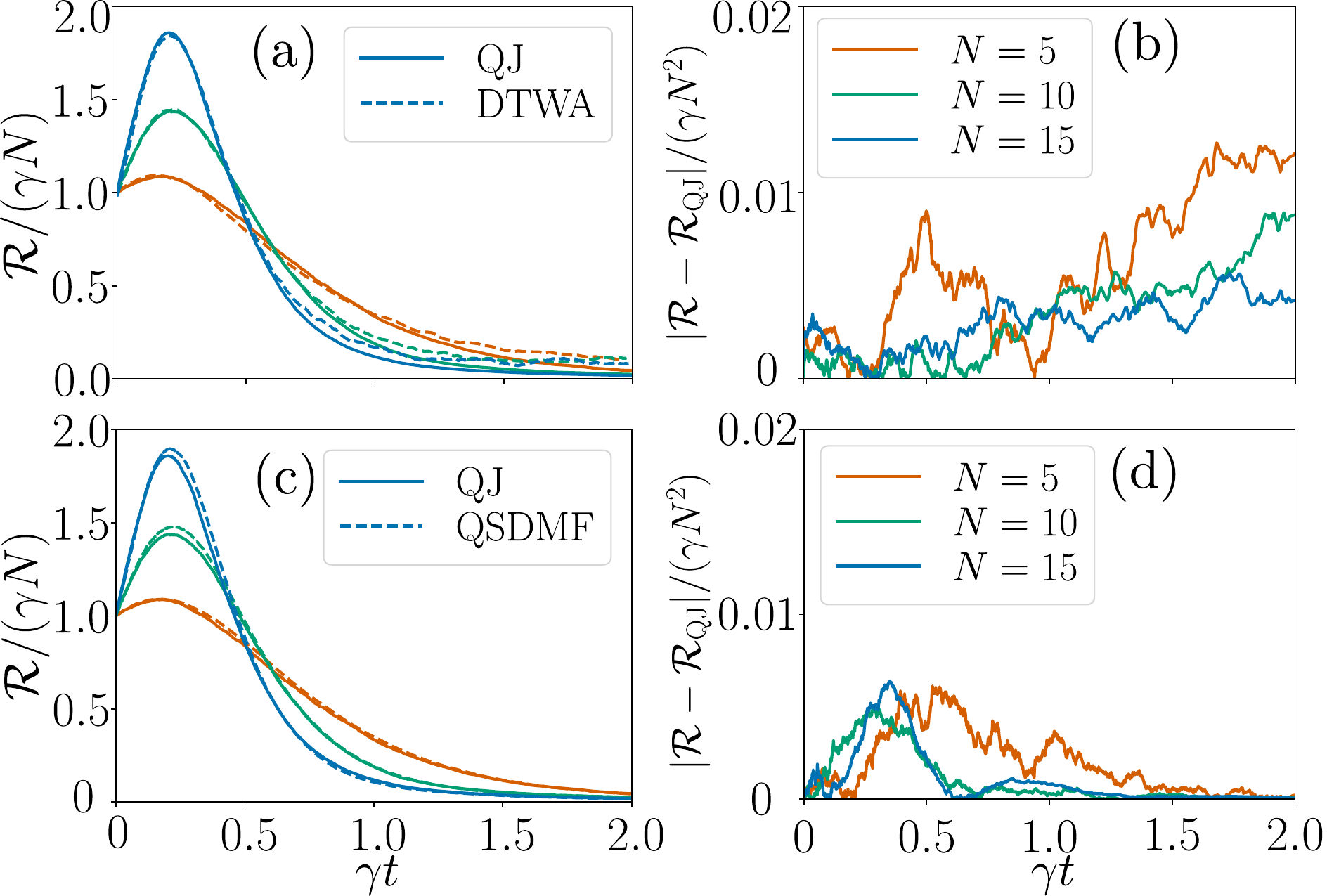}
	\caption{Benchmarking of DTWA and QSDMF methods. (a) and (c)  Plot of the collective decay rate $\mathcal{R}(t)$ for $N = 5, 10, 15$ (bottom to top) and for a disorder strength of $\Theta = \pi$. The results obtained from exact quantum jump (QJ) simulations (solid lines) are compared with those from DTWA and QSDMF simulations (dashed lines) with $\mathcal{N}=10^{3}$ trajectories. Using the same methods, (b) and (d) show the corresponding error for the normalized decay rate $\mathcal{R}(t)/N^2$ when compared to QJ results. 
    }
	\label{FigQT}
\end{figure}

In Fig.~\ref{FigQT}, we benchmark both numerical methods against exact quantum jump simulations for small atom numbers, $N=5, 10, 15$, and for a strongly disordered atomic array. The quantity of interest is the collective decay rate $\mathcal{R}(t)$, which is computed as
\begin{align}\label{EqR}
    \mathcal{R}(t) &= \,\frac{\gamma}{2}\braket{ J_{\text{R}}^{\dagger} J_{\text{R}}} + \frac{\gamma}{2}\braket{ J_{\text{L}}^{\dagger} J_{\text{L}} } \nonumber \\
      &\approx  \sum_{j,l\neq j} \frac{\gamma}{4} \cos(\xi_{j}-\xi_{l})  \overline{(s_{j}^{x}s_{l}^{x} + s_{j}^{y}s_{l}^{y})} +  \sum_{j} \frac{\gamma}{2}  ( 1 + \overline{s_{j}^{z}} ),
\end{align}
where $\overline{\bullet}$ denotes averaging over semiclassical trajectories that are obtained either from the DTWA or the QSDMF simulations.

The comparison shows excellent agreement between semiclassical simulations and quantum jump results, even for modest system sizes, as well as a systematic reduction of the relative errors for increasing $N$. This concerns, in particular, the location and height of the superradiant peak, which are the focus of our discussion below. For DTWA, the comparison becomes less accurate at the final stage of the decay, which can be attributed to the MF-like factorization of the photon-atom coupling in Eq.~\eqref{SpinEOM} and Eq.~\eqref{AlphaSDE}. 
However, we conclude that for the transient regime of interest, both the DTWA and the QSDMF simulations capture the collective decay dynamics very accurately, while being at the same time numerically highly efficient and scalable to hundreds or thousands of atoms. In addition, similar to Ref.~\cite{ZhangPRL2025}, an argument for the underlying reason behind the accuracy of QSDMF for large $N$ is presented in Appendix \ref{AppdxMutualInfo}.

\subsection{Comparison with Other Methods}\label{secMethodsCompareAnalysis}

Beyond the two methods presented above, there are other alternative methods that are in principle suited for the simulation of superradiance, and for comparison, additional simulation results are presented in Appendix~\ref{CompareMethods}.
First of all, a variant of the DTWA based on spin-length conserving trajectories~\cite{MinkSciPostPhys2023} permits a direct simulation of the collective spin dissipation. This method has a slightly better accuracy at the final stage of the decay, but otherwise the same accuracy and the same complexity $\mathcal{O}(N)$ as the method outlined above. Both methods produce unphysical trajectories, which means that only statistical averages correspond to physical observables. While this property restrains a simple physical interpretation of individual trajectories, their ensemble average can encode entanglement~\cite{SchachenmayerPRX2015}, which is not possible with QSDMF.

Methods based on a cumulant expansion (CE) of correlations \cite{Kubo1962,PlankensteinerQuantum2022,OriolPRR2023} scale as $\mathcal{O}(N^C)$, where $C \geq 2$ is the truncation order.  As a result, such simulations become costly for large systems. Both our DTWA approach and CE methods can handle non-Markovian effects, for example, by coupling the atoms to additional cavity modes. However, as shown in Sec.~\ref{secNonMarkov} and Appendix~\ref{CompareMethods}, in this non-Markovian setting, CE requires $C>3$ to converge, which increases the computational cost significantly. On the other hand, our DTWA method is still very accurate while keeping $\mathcal{O}(N)$ complexity.   

To conclude, out of these various methods,  the DTWA approach described above combines an efficient scaling \footnote{Note that, for ease of comparison, the scalings here are for short-range interactions. For long-range interactions, the scalings are modified by additional $N$-dependent factors. In the all-to-all case considered here, DTWA, QSDMF, and CE each pick up an extra factor of $N$, whereas QSD+MPS picks up an extra factor of $N^2$.}, $\mathcal{O}(N)$, with the ability to account for both classical and---up to a certain degree---quantum correlations and permits a straightforward and numerically inexpensive generalization to non-Markovian settings. This makes this method suitable for studying superradiance in a large variety of different scenarios, and has also been selected as the method of choice for most parts of the numerical analysis below. Additionally, we use QSDMF to gain trajectory-level insights, for example, on the spontaneous spin ordering studied in Sec.~\ref{SecUpperBoundR}.

\section{Superradiant Scaling in Disordered Arrays}\label{secNumericalResults}
Based on the numerical methods introduced above,  we now perform a systematic investigation of the scaling behavior of superradiance for a varying level of disorder and setting $z_{j} = \delta_{j}$~\footnote{While displacing the atomic positions by $z_j\rightarrow z_j+ m_j\times \lambda_0/2$ with $m_j\in\mathbbm{Z}$ does not affect the dissipators of our model, it can alter some terms of the Hamiltonian. However, this only shifts the precise values of $\mathcal{R}_\star/(\gamma N^2)$ by at most a few percent and leaves all qualitative features unchanged, as shown in both numerical simulations and the analytical analysis later. The results obtained for $z_j=\delta_j$ are thus representative for \emph{all} configurations where atoms are spaced by multiples of $\lambda_0/2$ on average.}. Of central interest in this study is the scaling of the peak emission rate $\mathcal{R}_\star$ and the burst time $t_{\star}$ with the number of atoms, $N$, and how this scaling is affected by disorder. In addition, we consider second-order photon-photon correlations, which are frequently used as an indicator for the onset of superradiance in the initial dynamics, but are less well understood during the subsequent burst dynamics.

\subsection{Superradiant Scaling}\label{secScalingNumerical}

\begin{figure}[t]
	\center
\includegraphics[width=\columnwidth]{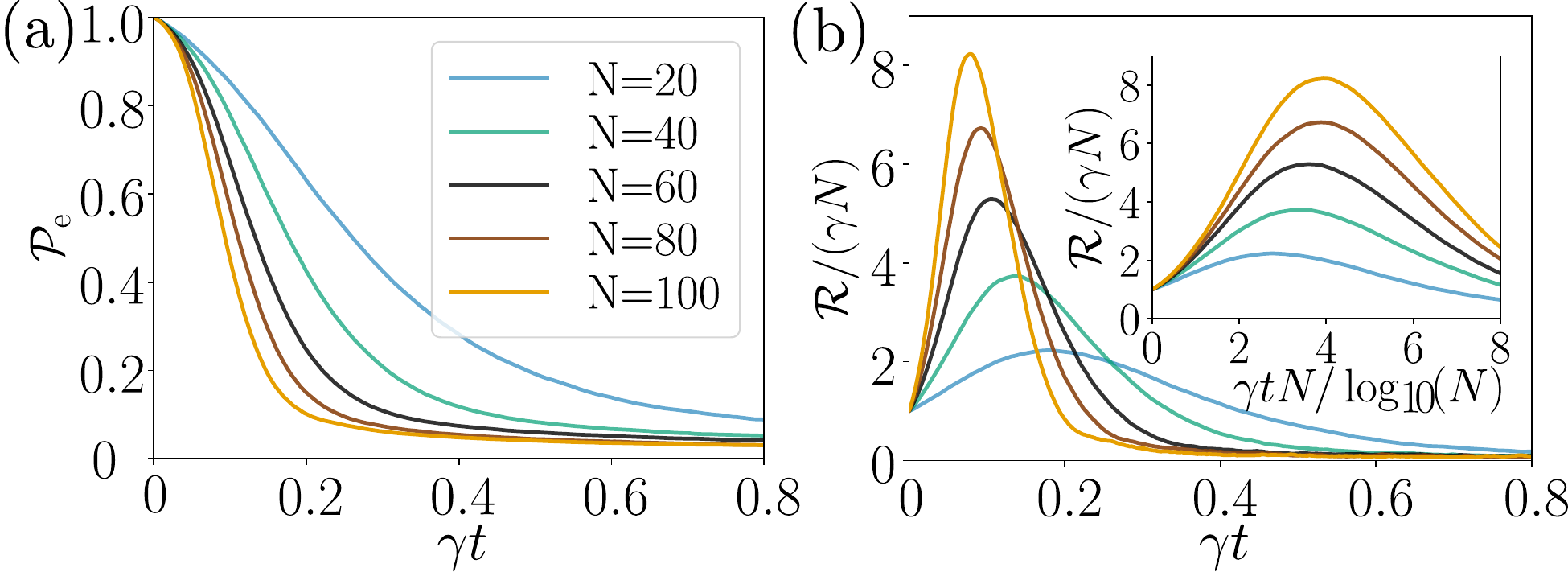}
	\caption{Superradiant dynamics for spatial disorder strength $\Theta = \pi$. (a) Decay of the average excited-state population $\mathcal{P}_{\rm e}$, showing an increasingly rapid decay as $N$ increases. (b) Plot of the corresponding decay rate per atom, $\mathcal{R}(t)/N$, which exhibits a sharp peak at the burst time $t_{\star} \sim \log(N)/N$. The inset shows the same data with a rescaled time. Both plots have been obtained using the DTWA method with averages taken over $\mathcal{N}=10^3$ trajectories.  
   }
	\label{FigDecayIllustration}
\end{figure}

As an illustrative example for disordered superradiance, we plot in Fig.~\ref{FigDecayIllustration} the disorder-averaged time evolution of the excited-state population, $\mathcal{P}_{\text{e}}=\sum_j \langle 1+\sigma^z_j\rangle/(2N)$, and the disorder-averaged decay rate $\mathcal{R}$ 
as a function of time and for a disorder strength of $\Theta = \pi$. Note that in our simulations, we choose one realization of the disordered locations per trajectory, such that the ensemble average gives the disorder-averaged density matrix and expectation values. From these results, we see an accelerated decay with a superradiance peak emerging at a burst time $t_{\star} \sim  \log(N)/N$, consistent with the behavior in the ideal Dicke model. The burst also sharpens and intensifies with increasing atom number $N$.

\begin{figure}[t]
	\center
\includegraphics[width=\columnwidth]{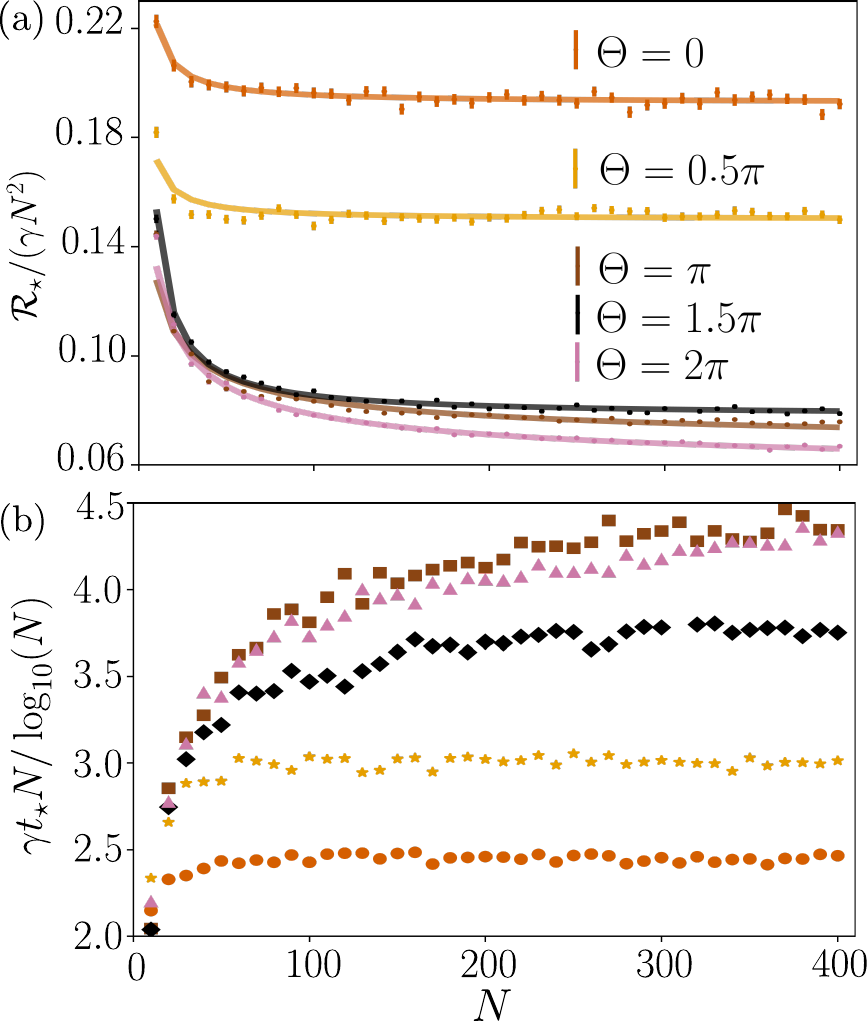}
	\caption{Superradiant scaling in the presence of spatial disorder. (a) Plot of the scaled superradiant rate, $\mathcal{R}_{\star}/(\gamma N^2)$, as a function of the atom number $N$ and for a varying disorder strength $\Theta$. The plot demonstrates the asymptotic recovery of the Dicke superradiant scaling with finite-size corrections that are fitted to the analytic prediction in  Eq.~\eqref{FiniteSizeFit} (solid lines). Similarly, the numerical data presented in (b) show the corresponding scaled burst time, $t_{\star} \gamma N/\log(N)$, which again converges to the asymptotic Dicke scaling for large $N$. In both plots, the results have been obtained using the DTWA method, averaged over $\mathcal{N}=10^3$ trajectories. 
    }
	\label{FigRandomPhaseDecay}
\end{figure}

To investigate this behavior more systematically, we evaluate the peak emission rate $\mathcal{R}_{\star} = \max_{t} \mathcal{R}(t)=: \mathcal{R}(t_{\star})$ and in Fig.~\ref{FigRandomPhaseDecay}(a) we plot the scaled value $\mathcal{R}_{\star}/(\gamma N^2)$ as a function of $N$ for different disorder strengths $\Theta$.  For all $\Theta$, the curves converge toward a constant value at large $N$, which confirms that the superradiant peak retains the ideal Dicke scaling $\mathcal{R}_{\star} \sim N^2$ asymptotically. As shown in Fig.~\ref{FigRandomPhaseDecay}(b), the same is true for the burst time $t_\star$, which also approaches the ideal Dicke scaling for very large $N$. 
Notably, however, for both quantities the convergence to the asymptotic scaling is considerably slower for large $\Theta$ and in particular for disorder strengths $\Theta = n\pi$, with $n \in \mathbb{Z}_+$. Overall, we find that the peak rates can be fitted using the scaling relations
\begin{equation}\label{FiniteSizeFit}
\mathcal{R}_{\star} =  
\begin{cases}
\gamma N^2 ( r_0 + r_1 N^{-1} ), & \text{if } \Theta \neq n \pi \\
\gamma N^2 ( r_0 + r_1^{\prime} N^{-1/2} ), & \text{if } \Theta = n \pi,
\end{cases}
\end{equation}
meaning that the rescaled maximal emission rate approaches the (disorder-dependent) numerical value of $r_0$ for very large $N$. The dependence of $r_0$ on the disorder strength is plotted in Fig.~\ref{bound} below, and we observe that $r_0$ converges to a constant value of about $r_0\approx 0.06$ for large disorder. This value is only reduced by about one-fourth compared to the fully ordered array, indicating that collective emission is rather robust even under strong disorder.

\subsection{Finite-Size Scaling}

To explain the different scalings toward the asymptotic limit, we return to the diagonalized form of the decay channels given in Eq.~\eqref{TwoChannelLindbladian}. Further, we make the simplifying assumption that, for a given disorder realization, the decay dynamics can be qualitatively captured by the maximal eigenvalue of the decay matrix, $\Gamma_+=\gamma N(1+ |c|)/2$, which is determined by the overlap parameter in Eq.~\eqref{cDef}. For phases $\xi_j$ that are uniformly distributed in $[-\Theta/2, \Theta/2)$, the mean value $\mu_a$ ($\mu_b$) and the variance $\sigma_a^2$ ($\sigma_b^2$) for the quantity $ \Re e^{i2\xi_j}$ ($ \Im e^{i2\xi_j} $) are given by
\begin{align}
\mu_{a} & = \frac{\sin{\Theta}}{\Theta} ,  & \sigma_{a}^{2} &=  \frac{( \Theta^2+ \cos{2\Theta} + \Theta \cos{\Theta} \sin{\Theta} -1)}{2\Theta^2}, \nonumber \\
\mu_{b} & =0, &  \sigma_{b}^{2} &=  \frac{( \Theta - \cos{\Theta}\sin{\Theta} )}{2\Theta}.  \label{MeanVar}
\end{align}
We identify two qualitatively distinct cases. When $\mu_a\neq 0$ and $N\to\infty$, the central limit theorem states that $c$ is a Gaussian variable with $\Re c \sim N(\mu_{a}, \sigma_{a}^2/N)$ and $\Im c \sim N(\mu_{b}, \sigma_{b}^2/N)$. Thus, we obtain the disorder-averaged rate 
\begin{equation} \label{NSScaling}
 E\left[ \Gamma_{+}\right] \simeq  \frac{1}{2} \gamma N \left( 1 + |\mu_{a}| \right) = \frac{1}{2} \gamma N \left( 1 + \left|\frac{\sin\Theta}{\Theta}\right|  \right) , 
\end{equation}
with corrections to $ E\left[ \Gamma_{+}\right] /(\gamma N)$ scaling as $1/N$ (see Appendix \ref{SecSingleExcitationUpperBound}). In the limit $\mu_a=0$, which is the case for $\Theta=n\pi$, the phase factors $e^{i2\xi_j}$ have a vanishing average and are dominated by fluctuations. 
In this case, we have $\sigma_a^2=\sigma_b^2=1/2$,
$E[|c|] = \sqrt{\pi/(4N)}$, and thus 
\begin{equation} \label{NSScalingThetaNPi}
E\left[ \Gamma_{+}\right] = \frac{1}{2} \gamma N \left( 1 + \sqrt{\frac{\pi}{4N}} \right), 
\end{equation}
in the limit of large $N$. This gives a different finite-size correction $\sim N^{-1/2}$. We remark that this situation already occurs for $\Theta=\pi$, where for the situation with regular lattice spacing $\lambda_0$, the atoms are still spaced by at least $\lambda_0/2$. The reason is that this remaining gap corresponds to a propagation phase of $k_0\lambda_0/2=\pi$ between neighboring atoms. Then this phase $\pi$ does not change the maximal decay rate $\Gamma_{+}$, since $c$ depends on $2\xi_{j}$.

While this analysis does not allow us to predict the numerical prefactor $r_0$, it shows that the observed scaling relations for the maximal decay rate during the burst can be obtained from the relation $\mathcal{R}_\star \propto N \Gamma_+$. This observation is rather surprising, as in view of the nonlinear dynamics or the presence of additional Hamiltonian terms, such a simple correspondence is a priori not expected. As discussed in Appendix \ref{AppdxOtherBounds}, this observation is, however, consistent with an upper bound that is proportional to $\Gamma_{+} N$.

Note that for any $\mu_a\neq 0$, the anomalous finite-size scaling $\sim 1/\sqrt{N}$ is observed up to an atom number of $N \lesssim (\sigma_a^2+\sigma_b^2)/\mu_a^2$, before there is a transition to a $1/N$ scaling for very large $N$. This explains the rather slow convergence for all disorder strengths $\Theta \gg1$. In Appendix \ref{AppdxGaussian}, we also present $\mathcal{R}_\star$ for an ensemble of atoms with Gaussian-distributed phases $\xi_j$. Again, we find a disorder-dependent scaling toward the asymptotic limit, which can be explained by this relation between the mean value and the fluctuations in the propagation phases.

\subsection{Second-Order Correlations}\label{Secg2}

\begin{figure}[t]
	\center
\includegraphics[width=\columnwidth]{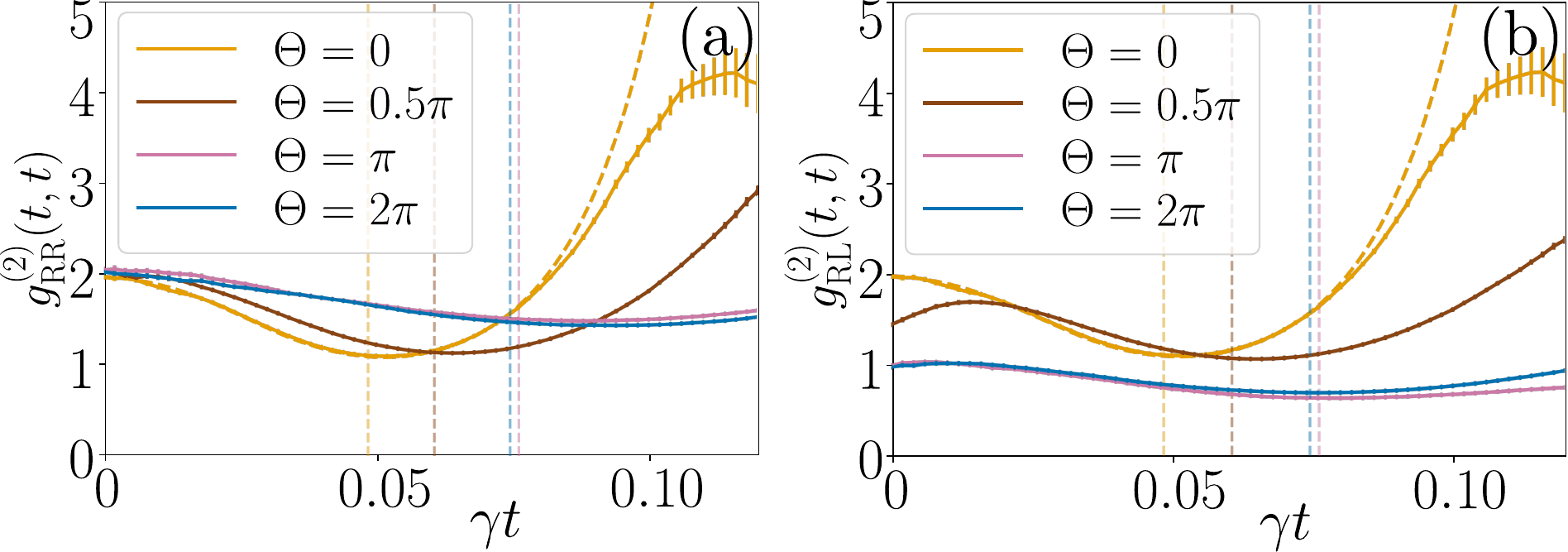}
	\caption{Photon-photon correlations. Plot of (a) the auto-correlations $g^{(2)}_{\text{R},\text{R}}(t,t)$ and (b) the cross-correlation $g^{(2)}_{\text{R},\text{L}}(t,t)$ during the superradiant decay for $\Theta = 0,\pi/2,\pi,2\pi$ and $N = 100$. The vertical lines indicate the burst time $t_{\star}$. The numerical results (solid lines) have been obtained using the DTWA approach with $\mathcal{N}=4000$ trajectories. For $\Theta = 0$, exact results obtained from rate-equation calculations are shown as a benchmark (dashed curve). 
    }
	\label{g2}
\end{figure}

Apart from the intensity of the emitted light, which is directly proportional to the emission rate, we can also study the second-order correlations of the emitted photons \cite{MassonNatComm2022,TebbenjohannsPRA2024,BachArXiv2024}. 
Specifically, we consider the equal-time second-order correlation functions defined as
\begin{equation}
    g^{(2)}_{\eta, \eta^\prime} (t,t) = \frac{ \braket{  J_{\eta}^{\dagger} (t) J_{\eta^{\prime}}^{\dagger}(t) J_{\eta^{\prime}}(t) J_{\eta}(t)  } }{ \braket{  J_{\eta^{\prime}}^{\dagger}(t) J_{\eta^{\prime}}(t)  }   \braket{  J_{\eta}^{\dagger}(t) J_{\eta}(t)  } },
\end{equation}
with $\eta, \eta^\prime \in \{ \text{R},\text{L}  \}$ denoting the two output channels. A correlation of $g^{(2)}_{\eta, \eta^\prime} (t,t)>1$ implies that it is more likely that a photon is emitted in the direction $\eta^\prime$ given that a previous photon has just been emitted into direction $\eta$.  

At the initial time \( t = 0 \), the autocorrelations are equal and given by 
\begin{equation}
\begin{split}
    g^{(2)}_{\eta,\eta} (0,0) &= \frac{ \sum_{j_1,j_2,j_3,j_4} \braket{ \sigma_{j_1}^{+} \sigma_{j_2}^{+} \sigma_{j_3}^{-} \sigma_{j_4}^{-} }_0  }{ \left( \sum_{j_1,j_2} \braket{ \sigma_{j_1}^{+} \sigma_{j_2}^{-} }_0\right)^2 } \\
    &=  \frac{2(N-1)}{N} \rightarrow 2,
\end{split}
\end{equation}
which is independent of the disorder realization.
Instead, the cross-correlations,
\begin{equation}\label{g2CrossCorr}
\begin{split}
    &g^{(2)}_{\text{R},\text{L}} (0,0) = g^{(2)}_{\text{L},\text{R}} (0,0) \\&= \frac{ \sum_{j_1,j_2,j_3,j_4} \braket{ \sigma_{j_1}^{+} \sigma_{j_2}^{+} \sigma_{j_3}^{-} \sigma_{j_4}^{-} e^{i  (\xi_{j_1}-\xi_{j_2}+\xi_{j_3}-\xi_{j_4} )}  }_0 }{ \left( \sum_{j_1,j_2} \braket{ \sigma_{j_1}^{+} \sigma_{j_2}^{-} }_0\right)^2 } \\
    &=\frac{N-1}{N}   \left[1 + \left(\frac{\sin \Theta}{\Theta}\right)^2 \right] \rightarrow 1 + \left(\frac{\sin \Theta}{\Theta}\right)^2,
\end{split}
\end{equation}
depend on the locations of the atoms and are reduced on average even in the initial stage [averaging is performed in the third line of Eq.~\eqref{g2CrossCorr}]. Specifically, this result shows that the initial decay into the two decay channels $\text{L}$ and $\text{R}$ is uncorrelated for a disorder strength that is an integer multiple of $\pi$ and in the limit of a uniform distribution, $\Theta\rightarrow \infty$.

For later times ($t\neq 0$), we compute $ g^{(2)}_{\eta,\eta^{\prime}} (t,t)$ numerically. The results are shown in Fig.~\ref{g2}. In the absence of disorder, both auto- and cross-correlations dip to a value of $\approx 1$ near the superradiant peak, as the light emitted is essentially coherent. Apart from a slight overall delay, the same behavior is still observed for moderate levels of disorder, despite a significant reduction of the cross-correlations at the initial stage. For strong disorder, we observe a different behavior. The dip in the autocorrelations is much less pronounced, and cross-correlations remain close to a value of $\approx 1$ during the whole decay process. The fact that the auto-correlations are consistently larger than cross-correlations implies an asymmetry in emission: a photon detected in one output channel is more likely to be followed by another photon in the same direction. This correlation has been recently studied in Ref.~\cite{CardenasLopezPRL2023} for small periodic atomic arrays.  As shown later, the persistence of this asymmetry becomes transparent and more evident when considering the heterodyne trajectories.

In addition to auto- and cross-correlations for the two channels, we can introduce the  photon-photon correlation function for the total emission,
\begin{equation}
    \tilde g^{(2)} (t,t) = \frac{\gamma^2}{4} \sum_{\eta,\eta^\prime=R,L} \frac{ \braket{  J_{\eta}^{\dagger} (t) J_{\eta^{\prime}}^{\dagger}(t) J_{\eta^{\prime}}(t) J_{\eta}(t)  } }{\mathcal{R}^2(t) },
\end{equation}
to quantify the total probability of emitting two successive photons into any direction.  An initial value of $\tilde g^{(2)} (0,0)>1$ implies that the emission of the first photon enhances the emission probability of the second photon compared to two completely independent emission events. Consequently, the decay rate $\mathcal{R}$ initially increases as a function of time and reaches a maximum during the subsequent decay. The condition $\tilde g^{(2)} (0,0)>1$  has thus been proposed as a signature of superradiance~\cite{MassonNatComm2022}, and more recently used to conjecture the superradiant scaling $\mathcal{R}_{\star} \sim N^{\tilde{g}^{(2)}}$~\cite{HolzingerArxiv2025}.

For our 1D atomic array, the disorder-averaged total correlation function at the initial time is given by 
\begin{equation}
    \tilde g^{(2)} (0,0) =\frac{N-1}{N}   \left[\frac{3}{2} + \frac{1}{2}\left(\frac{\sin \Theta}{\Theta}\right)^2 \right] \rightarrow \frac{3}{2} + \frac{1}{2}\left(\frac{\sin \Theta}{\Theta}\right)^2,
 \end{equation}
and, for sufficiently large $N$, it approaches a value of $\tilde g^{(2)} (0,0) \geq 3/2$ for any level of disorder. Together with a full simulation of $\mathcal{R}$, this observation confirms the correspondence between photon bunching correlations at the initial stage and the existence of a true superradiant burst at a later time. However, our simulations of the maximal decay rate $\mathcal{R}_\star$ show that, beyond finite-size effects, there is no additional influence of the initial photon-photon correlations on the superradiant scaling relations and that $\mathcal{R}_\star\sim N^2$ and $t_{\star}\sim \log(N)/N$ are established independently of the value of $\tilde g^{(2)} (0,0)$.

Finally, let us remark that at times $t> t_{\star} $, the photon emission statistics becomes more and more dominated by rare events, where the superradiant burst is delayed compared to the average decay, and by population trapped in slowly-decaying subradiant states. This leads to divergent photon-photon correlations, which are no longer accurately captured by DTWA simulations, but do not affect any of the conclusions from above about the superradiant peak.


\section{Origin of robust superradiance} \label{SecUpperBoundR}

The numerical observation that not only the phenomenon of superradiance but also the associated scaling relations are robust against strong spatial disorder is rather surprising, since for any spatial arrangement, the atomic dipoles interfere both constructively and destructively. To gain insight into how superradiance can persist under such conditions, we present in this section additional analytic estimates and upper bounds for the collective decay rate, based on a product-state ansatz for the atomic states. This analysis also reveals a spontaneous spin-ordering mechanism that underlies the robustness against spatial disorder.    


\subsection{Maximal Decay Rate}\label{secProductAnsatz}

\begin{figure}[t]
	\center
\includegraphics[width=0.8\columnwidth]{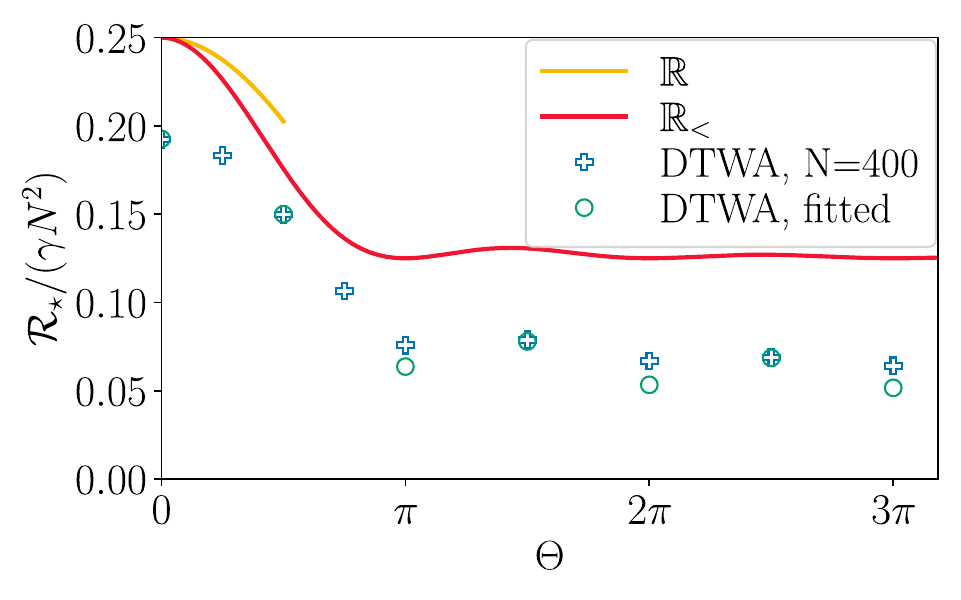}
	\caption{Asymptotic limit. Plot of the scaled superradiant rate $\mathcal{R}_{\star}/(\gamma N^2)$ for large atom numbers $N$ and as a function of the disorder strength $\Theta$. Circles show the fitted values of $r_0$ in Eq.~\eqref{FiniteSizeFit}, while crosses represent the rates for a large, but finite $N = 400$. The solid lines show (i) the exact analytic bound derived in Eq.~\eqref{ExactUpperBound} for $\Theta < \pi/2$  and (ii) the approximate upper bound $\mathbb{R}_{<}$ given in Eq.~\eqref{Rstarlower}. The qualitative agreement between the numerical data and $\mathbb{R}_{<}$ obtained from a product state ansatz confirms the semiclassical nature of superradiance and that the asymptotic $N^2$-scaling remains intact at arbitrarily large disorder strength.}
	\label{bound}
\end{figure}

While superradiance was originally described in terms of transitions between highly entangled collective states~\cite{DickePhysRev1954}, most features of this phenomenon can be explained in terms of classical correlations between the atomic dipoles. More specifically, it has been shown~\cite{ZhangPRL2025} that in general, when $N$ is large, the superradiant decay can be accurately described by a mixed state of the form (see also Appendix \ref{AppdxMutualInfo}),
\begin{equation}\label{eq:ProductState}
    \rho = \sum_{s} p_{s} \ket{\psi^{(s)}_{\text{prod}}}\! \bra{\psi^{(s)}_{\text{prod}}}.
\end{equation}
Here, the individual contributions,
\begin{equation}\label{ProductState}
    \ket{\psi_{\text{prod}}} = \bigotimes_{j} \left(\cos\frac{\theta_{j}}{2} \ket{e}_j + e^{i \phi_{j}} \sin\frac{\theta_{j}}{2} \ket{g}_j\right),
\end{equation}
are products of spin-coherent states, and represent sets of classical dipoles that are oriented along the polar angles $\theta_j$ and $\phi_j$. 

For a given product state, the corresponding decay rate is given by 
\begin{equation}\label{Rlower}
\begin{split}
&\frac{\gamma}{2} \braket{\psi_{\text{prod}} | \left( J_{\rm R}^\dag J_{\rm R}+J_{\rm L}^\dag J_{\rm L} \right) |\psi_{\text{prod}}}  \\
    &= \frac{\gamma}{4} \sum_{i,j} \cos( \xi_{i} - \xi_{j} )  \cos(\phi_{i} - \phi_{j})  \sin\theta_{i} \sin\theta_{j}.
\end{split}
\end{equation}
By maximizing this expression with respect to all angles $\{ \theta_{j}, \phi_{j}\}$, we obtain the upper bound $\mathbb{R}$, which---under the validity of Eq.~\eqref{eq:ProductState}---represents the maximal emission rate that can possibly be reached for a given disorder configuration. Since this maximum is always reached for $\theta_{j} = \pi/2$, i.e., when the radiating dipoles point along the XY-plane of the Bloch sphere, we find
\begin{equation}\label{OurBound}
\begin{split}
\mathcal{R}_{\star} \leq \mathbb{R} = \max_{ \{ \phi_{j} \} } \sum_{i,j} \frac{\gamma}{4} \cos( \xi_{i} - \xi_{j} )  \cos( \phi_{i} - \phi_{j} ).
\end{split}
\end{equation}
For a general atomic configuration, it is not possible to perform this maximization in a closed form.

Recently, by applying a theorem proved by Lieb \cite{lieb_classical_1973} and  Bravyi {\it et al.} \cite{bravyi_approximation_2019} for bounding the ground state energy of spin-$1/2$ Hamiltonians, an upper bound for $\mathcal{R}_{\star}$ has been derived in Ref.~\cite{MokArxiv2024}. By refining this calculation, and after taking the disorder average, we obtain (see Appendix \ref{AppdxOtherBounds})
 \begin{equation}
\begin{split}
\mathcal{R}_{\star} \leq \mathbb{R} \leq   \frac{3\gamma}{4} N^2 \left( 1 + \left|\frac{\sin\Theta}{\Theta}\right|  \right),
\end{split}
\end{equation}
in the limit of large $N$. While consistent with the $N^2$-scaling, we see that for $\Theta=0$ this bound overestimates the known scaling of Dicke superradiance by more than a factor of six. It is therefore not particularly useful to obtain more accurate predictions for the numerical prefactor $r_0$, nor does it provide further insights into the actual dipole configurations that optimize the decay rate. The reason for this overestimation lies in the fact that it ignores the approximately separable nature of superradiance.

Instead, the bound in Eq.~\eqref{OurBound} arises from an optimization over separable states, which turns out to be a much tighter bound. We first note that for weak disorder with $\Theta \leq \pi/2$, the quantities $\cos( \xi_{i} - \xi_{j} ) \geq 0$ are all positive. This makes $\{ \phi_{i} = \phi_{j} \}$ the optimal configuration and we obtain the exact upper bound 
\begin{equation}
\begin{split}\label{ExactUpperBound}
    \mathbb{R} &= \frac{\gamma}{4}  \sum_{i,j} E\left[ \cos( \xi_{i} - \xi_{j} )\right] \rightarrow \frac{\gamma}{4} N^2 \left[\frac{2}{\Theta} \sin\left(\frac{\Theta}{2}\right) \right]^2,
\end{split}
\end{equation}
which is also very close to the numerical values for $\mathcal{R}_\star$ plotted in Fig.\,\ref{bound}. Note that already in the absence of disorder, the rate $\mathcal{R}_\star/(\gamma N^2)\simeq 0.19$ is slightly lower than the maximally possible rate of $\mathbb{R}/(\gamma N^2)=1/4$, which can be attributed to the fact that the state in Eq.~\eqref{eq:ProductState} represents a mixture of trajectories, for which the superradiant burst occurs at slightly different times \cite{GrossPhysRep1982} (see also \cite{ZhangPRL2025} for the mixture of QSDMF trajectories). Leaving this averaging effect aside, this comparison suggests that even in the presence of disorder, the dipole tends to maximize the decay rate during the superradiant burst and that we can use $ \mathbb{R} \gtrsim \mathcal{R}_{\star} $ as a rather tight upper bound.

In general, it is not possible to evaluate $\mathbb{R}$ explicitly or to extend the result in Eq.~\eqref{ExactUpperBound} to larger values of $\Theta$. We notice, however, that the choice $\{|\phi_{i} -\phi_{j}|= |\xi_{i}-\xi_{j}|\}$ ensures that all contributions to the sum in Eq.\,\eqref{ExactUpperBound}  are positive such that 
\begin{equation} \label{RstarlowerDef}
   \mathbb{R}  \geq   \frac{\gamma}{4} \sum_{i,j}  \cos^2(\xi_{i}-\xi_{j} ) =: \mathbb{R}_{<}.
\end{equation}
Therefore, by taking the disorder average, we obtain a lower estimate 
\begin{equation}\label{Rstarlower}
\mathbb{R}_{<} \rightarrow \frac{\gamma}{8} N^2 \left[1 + \left(\frac{\sin \Theta}{\Theta}\right)^2 \right].
\end{equation}
Under the assumption that $\mathbb{R}$ is smaller than the bound without disorder, $\gamma N^2/4$, we can constrain the maximal decay rate to a rather narrow interval, $\mathbb{R}_{<} \leq \mathbb{R} \leq \gamma N^2/4$, or $(\mathbb{R} - \mathbb{R}_{<})/(\gamma N^2) \leq 1/8$. While this is a constraint on $\mathbb{R}$ and not on $\mathcal{R}_{\star}$, we find from numerics that the correspondence between $\mathbb{R}_{<} $ and $\mathcal{R}_\star$ is very close. The discrepancy can be attributed to the distribution around the optimal configurations in both the azimuthal angles $\phi_{j}$ and the polar angles $\theta_{j}$ (see the Supplemental Material \cite{SupMat}), which is neglected in deriving the upper bound. Based on this observation we conjecture that the dipole configurations $\{|\phi_{i} -\phi_{j}|= |\xi_{i}-\xi_{j}|\}$ that enter $\mathbb{R}_{<}$ also play a dominant role in the actual dynamics. This conjecture is also supported by the basic observation that $\mathcal{R}_\star$ saturates above a value of $\Theta\gtrsim\pi$. At this value, the distribution of \emph{twice} the phase differences, $2(\xi_i - \xi_j) \bmod (2\pi)$,  which enters the approximated expression in Eq.~\eqref{RstarlowerDef}, becomes flat. Instead, for a decay rate that only depends on the actual propagation phases, $\Delta \xi_{ij} = (\xi_i - \xi_j) \bmod (2\pi)$, a saturation of $\mathcal{R}_\star$ would be expected for $\Theta\gtrsim 2\pi$ (see the Supplemental Material \cite{SupMat}).

Note that the upper bound $\mathbb{R}$ 
is independent of the Hamiltonian $H$, which indicates that details in $H$ do not change the superradiant scaling. For example, as shown in Appendix \ref{IgnoreAtomInteraction}, artificially ignoring the Hamiltonian only modifies the prefactor of the $N^2$-scaling.

\subsection{Spontaneous Spin Ordering}\label{SubSecSpinOrdering}

\begin{figure}[t]
	\center
\includegraphics[width=\columnwidth]{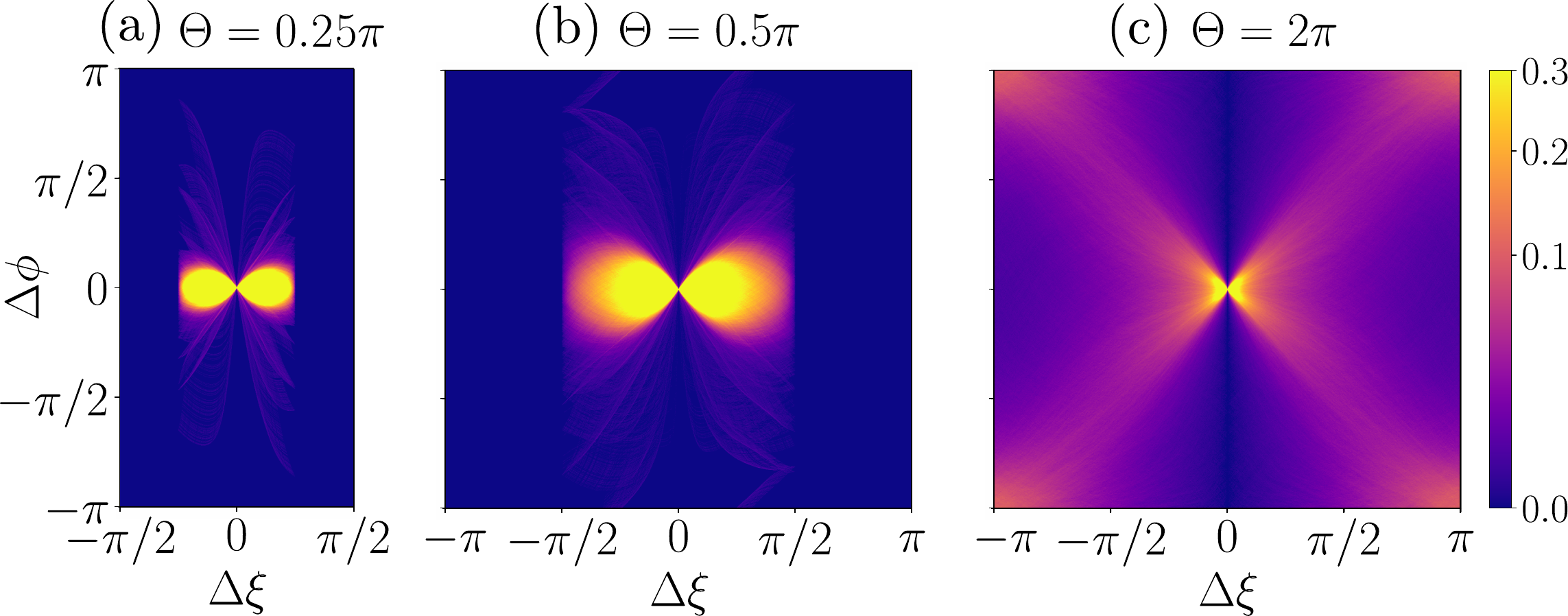}
	\caption{
    Spontaneous spin ordering. The joint probability distribution $p(\Delta \xi, \Delta \phi)$ of the relative propagating phase $\Delta \xi$ and relative dipole orientation $\Delta \phi$ is plotted for (a) $\Theta = \pi/4$, (b) $\Theta = \pi/2$,  and (c) $\Theta = 2\pi$. For strong disorder ($\Theta = 2\pi$), the distribution aligns along the diagonals $\Delta \phi \approx \pm \Delta \xi$, consistent with the predicted spin ordering pattern. For more data on the 
    crossover between weakly and strongly disordered configurations, see the Supplemental Material~\cite{SupMat}. Simulations are performed using the QSDMF method with $N=400$ atoms and $\mathcal{N}=10^3$ trajectories.}
	\label{FigSpinAlignment}
\end{figure}

In our analysis above, we find that alignments of dipoles satisfying $\{|\phi_{i} -\phi_{j}|= |\xi_{i}-\xi_{j}|\}$ lead to close-to-maximal decay rates for a given disorder realization. Therefore, we expect that these configurations will also dominate the actual mixed state near the time of the superradiant burst. Importantly, this condition is satisfied by two sets of configurations, $\{ \phi_{j} = \phi_{1} + (\xi_{j} - \xi_{1})  \}$ and  $\{ \phi_{j} = \phi_{1} - (\xi_{j} - \xi_{1})  \}$, which both fix the relative phases of the individual dipole according to their location along the waveguide [see Fig.~\ref{FigSchematic}(b)]. Because the system has a global $U(1)$ symmetry, the phase of the first atom, $\phi_1$, remains undetermined. 
By assuming that  only those configurations contribute, we can approximate the state at the burst time $t_{\star}$ as 
\begin{equation}\label{eq:Rhotau}
    \rho(t_{\star}) \approx   \int_0^{2\pi}  \frac{d\phi_{1} }{4\pi}    \left[ \rho^{+} \left(\phi_{1}\right) + \rho^{-} \left(\phi_{1}\right)  \right],
\end{equation}
where $\rho^{\pm} \left(\phi_{1}\right)$ represent distributions around the two product states with atomic dipoles that rotate clockwise (anticlockwise) according to the phase patterns identified above. Since both configurations result in identical decay rates, they are weighted equally. Therefore, apart from the usual $U(1)$ symmetry, the superradiant state exhibits an additional emergent $Z_{2}$ symmetry in the presence of strong disorder.

To test the validity of the prediction for $\rho(t_{\star})$ in Eq.~\eqref{eq:Rhotau}, we take a closer look at the spin orientations obtained from our QSDMF simulations. Specifically, for a given disorder realization, we evaluate all propagation phases $\Delta \xi_{ij} = (\xi_i - \xi_j) \bmod (2\pi)$ together with the set of relative dipole orientations, $\Delta \phi_{ij} = (\phi_i - \phi_j) \bmod (2\pi)$ in the QSDMF trajectories at $t=t_{\star}$. In Fig.~\ref{FigSpinAlignment}, we show the resulting joint probability density $p(\Delta\xi, \Delta\phi)$ for three different strengths of the disorder. According to our analysis above, for weak disorder we expect this distribution to be peaked around $\Delta \phi=0$, which maximizes the decay rate independently of the value of $\Delta \xi$. In contrast, for a strongly disordered system, the probability density should be peaked around the values $\Delta \phi=\pm \Delta \xi$, i.e., along the diagonals of the plots. 

Indeed, we find that both trends are clearly visible in the examples shown in Fig.~\ref{FigSpinAlignment}, although considerably broadened by fluctuations. This confirms that, under strong disorder, the spins self-organize their orientations along a local direction determined by their random position. This is in stark contrast to the weakly disordered cases, where spins tend to synchronize with a common phase.  Interestingly, under all conditions, we find a particularly strong alignment tendency for atoms that are multiples of a wavelength apart, i.e., $\Delta \xi=0$. In this case, there is an almost vanishing probability to find a nonzero relative phase $\Delta \phi$, between the corresponding dipoles. Note that these findings are not constrained to systems with disorder. In Appendix~\ref{SpinOrderingRegularLattice}, we perform an identical analysis for periodic arrays, where for lattice spacing that do not match the emission wavelength $\lambda_0$, similar orderings of the atomic dipoles arise.

Our finding of spontaneous spin ordering during superradiant decay is conceptually related to other works on synchronization in superradiance. In particular, Ref.~\cite{ZhuNJP2015} studied an array of quantum dipoles under continuous pumping, and showed that long-range interactions and collective decay can lead to synchronization to a common phase in the steady state. 
Instead, our result reveals a distinct dynamical mechanism: during spontaneous emission without external control, the spins self-organize their orientations transiently in a disorder-dependent pattern that enhances constructive interference and preserves superradiant scaling.

\subsection{Mirror-Asymmetric Correlations}

\begin{figure}[t]
	\center
\includegraphics[width=\columnwidth]{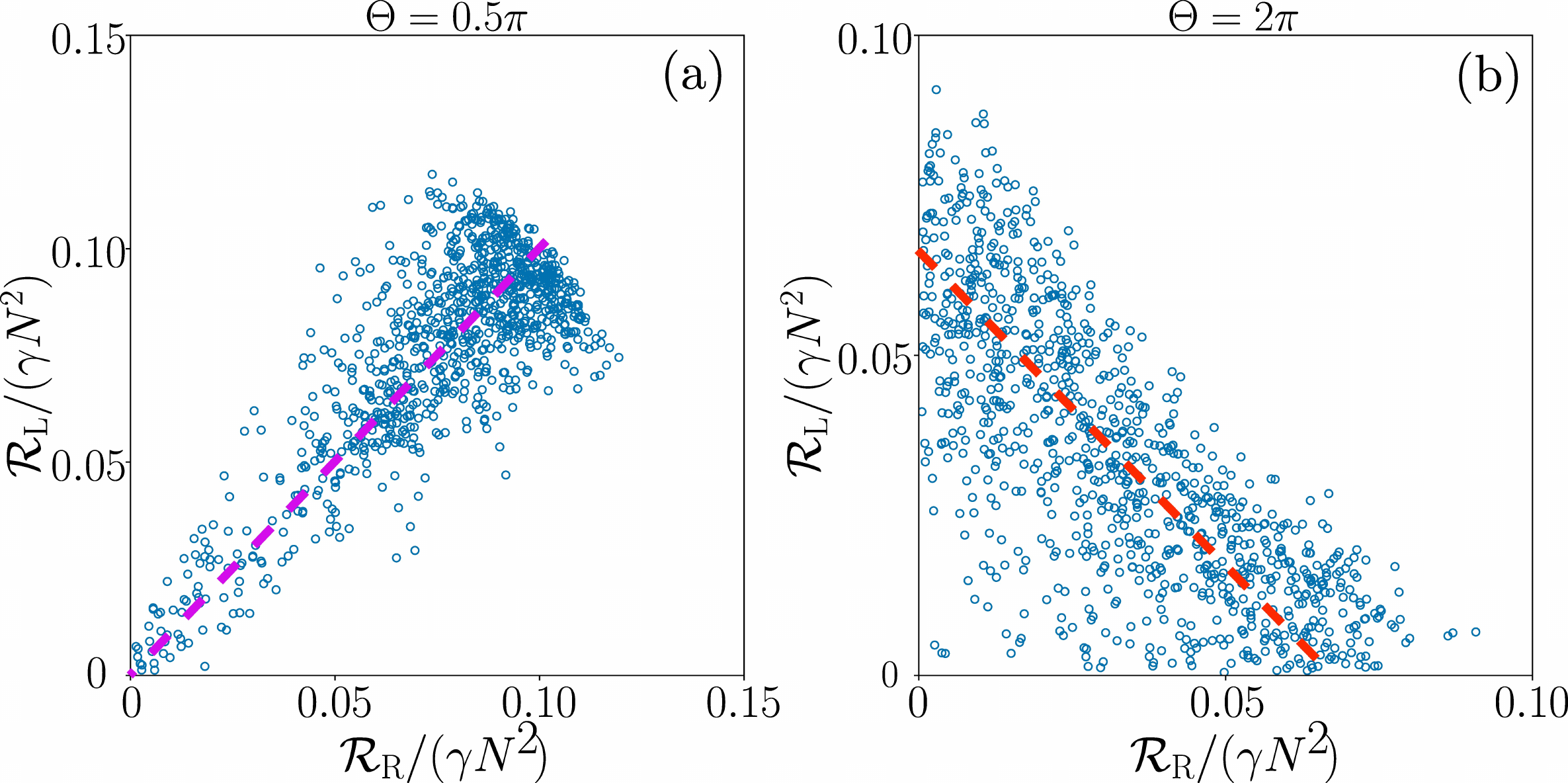}
	\caption{Asymmetric emission. Plot of the joint distribution of rates $(\mathcal{R}_{\text{R}},\mathcal{R}_{\text{L}})$ at the superradiant time $t_{\star}$, as obtained from QSDMF simulations with $N=400$, $\mathcal{N}=10^3$ trajectories and (a) $\Theta=\pi/2$ and (b) $\Theta=2\pi$. For weak disorder, photon emission to the right and the left occurs with similar rates, $\mathcal{R}_{\text{R}}\approx\mathcal{R}_{\text{L}}$ [as shown by the dashed line in (a)]. For large disorder, the decay rates are anti-correlated with $\mathcal{R}_{\text{R}} + \mathcal{R}_{\text{L}} \approx \mathcal{R}_{\star}$ [as shown by the dashed line in (b)].
    }
	\label{FigRRRL}
\end{figure}

In our analysis of photon correlation in Sec.~\ref{Secg2}, we have observed that for large disorder, the auto-correlations are enhanced compared to the cross-correlations. This condition implies that, while the emission of photons to the left and to the right is equally likely on average, two successive photons are roughly twice as likely to be emitted to the same side as to opposite directions. Similar correlations have previously been reported for small regularly-spaced arrays~\cite{CardenasLopezPRL2023}. We emphasize, however, that both in the system studied in Ref.~\cite{CardenasLopezPRL2023} and in the disordered atomic arrays considered here, the ratio between cross- and auto-correlation does not vanish, even in the limit of large $N$.

It turns out that this feature of the photon-photon correlations is connected to the spontaneous spin ordering discovered above: the configurations $\{ \phi_{j} = \phi_{1} + (\xi_{j} - \xi_{1})  \}$ and  $\{ \phi_{j} = \phi_{1} - (\xi_{j} - \xi_{1})  \}$ correspond to those configurations that maximize the emission into the left and into the right modes, respectively. More precisely,  by introducing the emission rates to the left and to the right, $\mathcal{R}_{\eta={\rm R,L}}=\frac{\gamma}{2}{\rm Tr}\{ J_\eta^\dag J_\eta \rho\}$, and considering product states as given in Eq.~\eqref{ProductState}, we obtain 
\begin{equation}
\begin{split}
    \mathcal{R}_{\text{R}} = \frac{\gamma}{8} \sum_{i,j\neq i} \sin\theta_{i} \sin\theta_{j} \cos( \phi_{i} -\phi_{j} + \xi_{i} - \xi_{j}  ), \\
    \mathcal{R}_{\text{L}} = \frac{\gamma}{8} \sum_{i,j\neq i} \sin\theta_{i} \sin\theta_{j} \cos( \phi_{i} -\phi_{j} - \xi_{i} + \xi_{j}  ). 
\end{split}   
\end{equation}
These rates are maximized for $\theta_{j}=\pi/2$ and the respective configurations from above, i.e., 
\begin{equation}
\left.  \mathcal{R}_{\rm L} \right|_{\rho^+} =  \left. \mathcal{R}_{\rm R} \right |_{\rho^-}  = \frac{\gamma }{8} N^2.
\end{equation} 
At the same time, the opposite rate, for example, 
\begin{equation}
\left.  \mathcal{R}_L \right|_{\rho^-} = \frac{\gamma }{8}\sum_{i,j\neq i} \cos\left[ 2 \left(\xi_{i} - \xi_{j} \right)  \right] \rightarrow \frac{\gamma N^2}{8}  \left(\frac{\sin \Theta}{\Theta}\right)^2,
\end{equation} 
is suppressed on average in the limit of large disorder. 

In Fig.~\ref{FigRRRL}, we plot the actual distribution of the pair of rates $(\mathcal{R}_{\text{R}},\mathcal{R}_{\text{L}})$ in each QSDMF trajectory at time $t=t_{\star}$. These plots provide a very intuitive way to illustrate the effect of mirror-asymmetric emission in superradiant decay. For weak disorder both rates are strongly correlated around $\mathcal{R}_{\text{R}}\approx\mathcal{R}_{\text{L}}$. This implies that, along each trajectory, photons are emitted approximately equally likely to the left and to the right. In the regime of strong disorder, the rates are anti-correlated with $\mathcal{R}_{\text{R}} + \mathcal{R}_{\text{L}}\approx \mathcal{R}_{\star}$. Therefore, in this regime, the atomic dipoles evolve into configurations that successively lead to preferred emission into either the right or the left. It is these successive emissions that imprint the location-specified phases onto the spins and lead to the self-organization of the two spin orderings. Additional simulations presented in Appendix~\ref{MoreData} show that this preference exists during most parts of the decay, although individual trajectories may also change their preferred emission direction during the process.

The anti-correlation between $\mathcal{R}_{\text{R}}$ and $\mathcal{R}_{\text{L}}$ stands in stark contrast to the theoretical predictions at the MF level without dynamical quantum fluctuations~\cite{GrossPhysRep1982,HaakePRA1979}. There, with fluctuations only in the initial condition, the left- and right-going emissions are predicted to be essentially symmetric and uncorrelated. This highlights the crucial role of the quantum fluctuations in our simulation methods, both in enabling accurate predictions and in revealing the underlying microscopic mechanism.

In summary, Fig.~\ref{FigSpinAlignment} and Fig.~\ref{FigRRRL} clearly highlight the difference in the spontaneously established ordering of the atomic dipoles in weakly and strongly disordered arrays. For strong disorder, the peaks in the diagonals of the distribution of $p(\Delta\xi, \Delta\phi)$ in Fig.~\ref{FigSpinAlignment}(c) and the anticorrelations seen in the rates in Fig.~\ref{FigRRRL}(b)  also demonstrate the relevance of those dipole orderings that maximize emission to the left or the right. However, as evident from the photon correlations and the distribution of spin orientations, these orderings are distributions with a considerable broadening around the optimal configurations, which is also the reason underlying the discrepancy between the upper bound $\mathbb{R}_{<}$ and actual numerical data shown in Fig.~\ref{bound}. Note that our analysis has focused on spin ordering in the bulk of the atomic array. In a related recent analysis of spin-ordering in pumped atomic arrays~\cite{CardenasLopez2025}, it was observed that atoms at the left (right) end align anticlockwise (clockwise) and therefore prefer emission to the left (right). In our setup, such boundary effects also exist, but their contribution to the overall dynamics and decay asymmetry is minor (see Appendix~\ref{BoundaryEffect}).

\section{Robustness of superradiance in general settings}\label{secGeneral}

Apart from disorders in the locations of the atoms, realistic systems also have imperfections in many other aspects, including, for example, inhomogeneous broadening of the transition frequency, disorders in atom-waveguide coupling, and non-Markovian effects that arise when the emission is mediated by a cavity mode or a narrow-band waveguide. With the help of the numerical methods introduced in Sec.~\ref{secNumerics}, the impact of all these imperfections can be studied systematically. In the following, we briefly discuss the case of frequency disorder and that of non-Markovian decay to illustrate further the robustness of superradiant scaling relations under various types of disorders.

\subsection{Inhomogeneous Broadening}\label{secDisorderedAtomFrequency}

We extend our previous model by assuming random fluctuations in the atomic frequencies. In this case, the frequencies of the individual atoms are chosen as $\omega_{j} = \omega_{0} + \delta \omega_{j}$, where the $\delta \omega_{j}$ are drawn from a Gaussian distribution with vanishing mean and a variance of $\Delta_\omega^2/4$. This can be included in our master equation in Eq.~\eqref{ME0} by adding the Hamiltonian 
\begin{equation}
    H_{\omega} = \sum_{j} \frac{\delta \omega_{j} }{2} \sigma_{j}^{z}.
\end{equation}
By changing into a rotating frame with respect to this Hamiltonian, we obtain the coupled set of EOM,
\begin{align}\label{SpinEOMHeisenberg}
    \frac{d}{dt} \braket{\sigma_{j}^{-}} &= - \frac{\gamma}{2} \braket{\sigma_{j}^{-}} + \frac{\gamma}{2} \sum_{l\neq j} \braket{\sigma_{j}^{z} \sigma_{l}^{-}} e^{i|\xi_{j} - \xi_{l}|} e^{i(\delta\omega_{j} - \delta\omega_{l}) t}, \nonumber \\
        \frac{d}{dt} \braket{\sigma_{j}^{z}} &= - \gamma \left(\braket{\sigma_{j}^{z}} +1\right) \\
        &\phantom{=}- \gamma \sum_{l\neq j} \left( \braket{\sigma_{j}^{+} \sigma_{l}^{-}} e^{i|\xi_{j} - \xi_{l}|} e^{i(\delta\omega_{j} - \delta\omega_{l}) t} + \text{c.c.} \right). \nonumber
\end{align}
From these equations, we already see that the frequency disorder introduces additional phases $|\delta\omega_{j} - \delta\omega_{l}| t $. However, since the superradiant decay occurs on an accelerated time-scale set by $t_\star \approx  \ln(N)/(N\gamma)$, the phases accumulated during the burst only scale as $ \Delta_\omega t_\star \sim (\Delta_\omega/\gamma) \ln N/N$ and become negligible in the limit of large $N$. 
This argument can be extended to higher cumulants of order $C$, for which additional phases of the order $\sim C (\Delta_\omega/\gamma) \ln N /N$ appear in the EOM. Therefore, under the assumption that the superradiant decay is accurately described by a cumulant expansion of finite order $C \ll (\gamma/\Delta_\omega) N/ \ln N $ (see Appendix \ref{CompareMethods}), random frequency fluctuations become irrelevant.

\begin{figure}[t]
	\center
\includegraphics[width=\columnwidth]{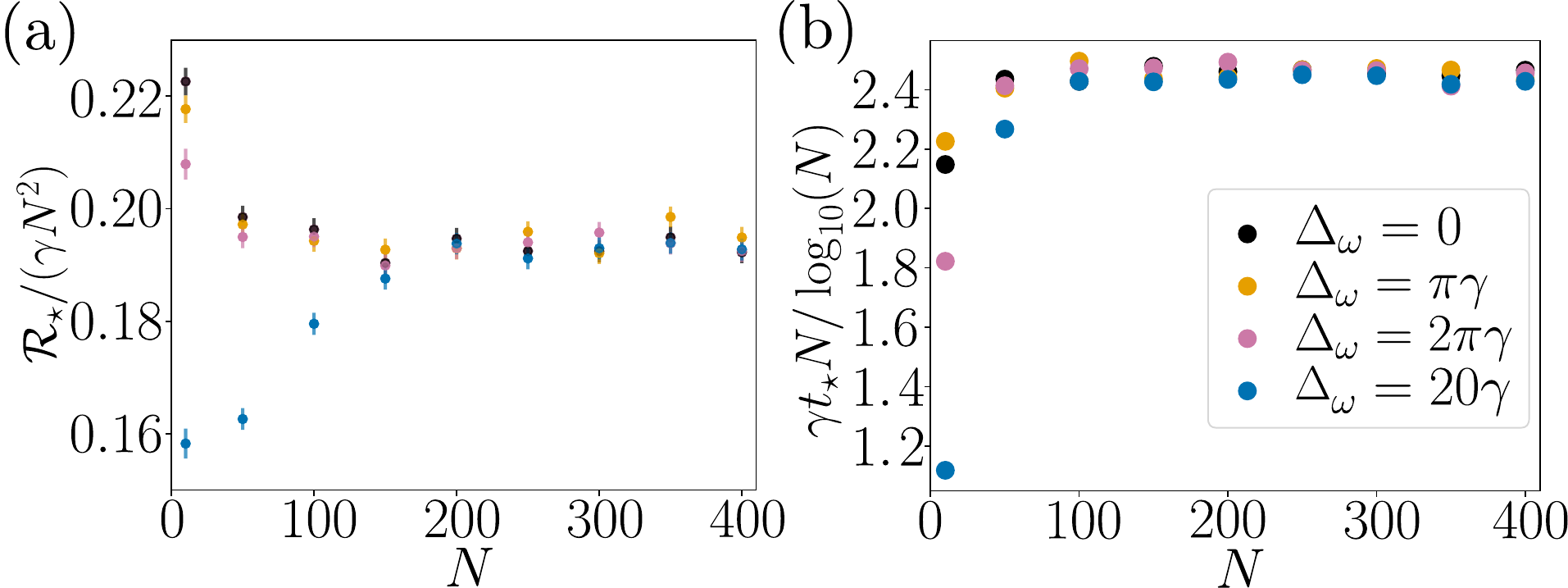}
	\caption{Superradiant scaling in the presence of frequency disorder. Plot of (a) the normalized superradiant rate, $\mathcal{R}_{\star}/(\gamma N^2)$, and (b) the rescaled burst time, $t_{\star}\gamma N/\log_{10}(N)$, as a function of the atom number $N$. The plotted results have been obtained using the DTWA method with $\xi_{j}=0$ and averaged over $\mathcal{N}=10^3$ trajectories. For each trajectory, the atomic frequency offsets $\delta \omega_j$ have been randomly drawn from a Gaussian distribution of width $\Delta\omega = 0, \pi\gamma, 2\pi\gamma, 20\gamma$. For the considered levels of disorder, the ideal Dicke scaling is recovered for sufficiently large $N$. }
	\label{FigDetuning}
\end{figure}

Indeed, in Fig.~\ref{FigDetuning}, we confirm this argument by simulating the effect of frequency disorder of increasing strength using the DTWA method. While there is a visible influence of inhomogeneous broadening when $N$ is moderately large, the deviations from the case without disorder vanish for sufficiently large $N$ even for very strong spectral disorder (see more detailed analysis in the Supplemental Material \cite{SupMat}).

\subsection{Non-Markovian Effects}\label{secNonMarkov}

Apart from various types of spatial and spectral disorders, another relevant question about superradiant decay concerns the role of non-Markovian effects. The non-Markovian regime has been notoriously difficult to study even for a few emitters \cite{BreuerRMP2016,deVegaRMP2017}, let alone large-scale simulations of $N$ emitters. Non-Markovian superradiance has only been investigated in very special cases, such as the cascaded system with time delays \cite{WindtPRL2025}, where an effective master equation can be derived \cite{CarmichaelPRL1992,GardinerPRL1992}. In the current setting, non-Markovianity can be explored in terms of the extended master equation introduced in Eq.~\eqref{extendedME}, where the collective decay channels are implemented via coherent coupling to two lossy cavity modes. In this setting, Markovianity and thus a recovery of the original master equation in Eq.~\eqref{ME0} are ensured when the dynamics of the atoms are slow compared to the decay times of the cavities, $\kappa^{-1}$. During the initial stage of the decay, this condition is satisfied when $\kappa \gtrsim \gamma N$. However, the situation is less clear around the time $t_\star$ of the superradiant burst, where, in view of the scaling $\mathcal{R}_\star\sim \gamma N^2$, the accelerated atomic decay may eventually be faster than the cavity loss. Specifically, under the condition $\gamma N \ll \kappa \ll \gamma N^2$, which can be encountered for sufficiently large $N$, the superradiant dynamics may both be slow and fast compared to the photonic environment, and the validity of a Markovian description is a priori unclear.

\begin{figure}[t]
	\center
\includegraphics[width=\columnwidth]{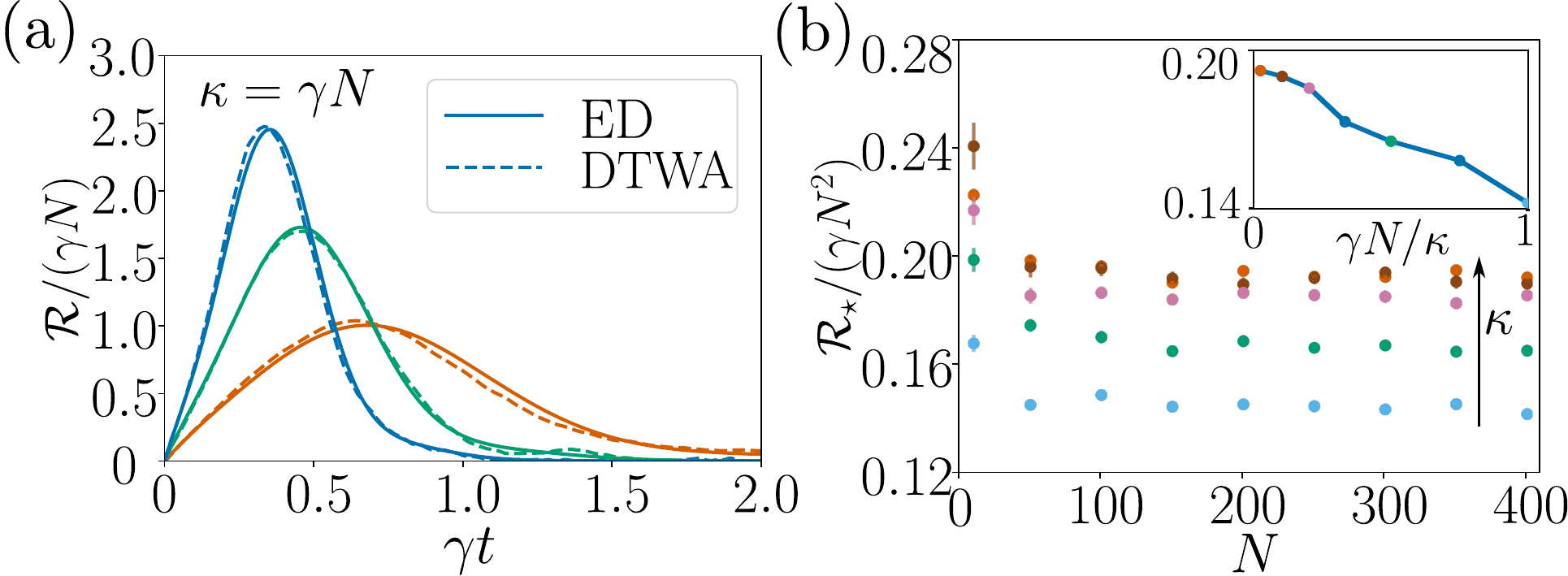}
	\caption{Impact of non-Markovianity on superradiant scaling. (a) Benchmark of the DTWA method against exact diagonalization (ED) of $N=5,10,15$ atoms coupled to one lossy cavity. (b) Scaling of the superradiant rate $\mathcal{R}_{\star}/(\gamma N^2)$ for $\kappa/(\gamma N) = 1, 2, 5, 10, 50$ (from bottom to top). The inset shows the value of $\mathcal{R}_{\star}/(\gamma N^2)$ for $N=400$ as a function of $\gamma N/\kappa$. Even in the non-Markovian regime [small values of $\kappa/(\gamma N)$], the $N^2$-scaling persists. The prefactor of $N^2$ scaling increases to the factor $\approx 0.2$ of the Dicke case as $\kappa$ increases. 
    All results have been obtained by averaging over $\mathcal{N}=10^3$ trajectories.
    }
	\label{FigNMDecay}
\end{figure}

In Fig.~\ref{FigNMDecay}, we use the full DTWA simulations described in Sec.~\ref{subsecDTWA} to study the combined evolution of the atoms and the cavity modes for varying ratios of $\gamma N/\kappa$ and as a function of $N$.  In Fig.~\ref{FigNMDecay}(a), we see the emergence of a clearly visible superradiant burst even for $\gamma N/\kappa=1$, i.e., at the very boundary of the regime for which non-Markovian behavior is expected. As shown in Fig.~\ref{FigNMDecay}(b) for this and any smaller ratio, the superradiant scalings $\mathcal{R}\sim \gamma N^2$ and $t_\star\sim \log(N)/(\gamma N)$ (see Supplemental Material \cite{SupMat}) are established for moderately high $N$, although the prefactor depends on the degree of non-Markovianity. Importantly, we find no deviation from these scaling relations even though the condition $\kappa \gg \gamma N^2$ is not met. 

These numerical results demonstrate that superradiance is robust against weak non-Markovianity, and clarify that the enhanced rate $\mathcal{R}_\star$ does not affect the validity of a Markovian description for superradiant decay as long as $\kappa \gg \gamma N$. 
We can reconcile this finding by noticing that the conditions for Markovianity from above are unnecessarily stringent, since the photons can leave the cavity modes at a rate $\sim \kappa N$ with $\mathcal{O}(N)$ photons inside. The relevant time scales are set by the atomic dynamics, which evolve on a timescale of order $1/\gamma$ initially and $1/(\gamma N)$ during superradiance. Consequently, it suffices to require $\kappa \gg \gamma$ initially and $\kappa \gg \gamma N$ in the superradiant regime to maintain Markovianity. Hence, the ratio $\gamma N/\kappa$ serves as a natural control parameter for the degree of non-Markovianity in the system.

\section{Conclusion and Outlook} \label{secConclusion}

\subsection{Summary}

In summary, we investigated the fate of Dicke superradiance in atomic arrays subject to strong disorder. Using semiclassical methods tailored for collective emission, we numerically demonstrate that the characteristic $N^2$-scaling of the peak emission rate and the scaling $t_\star\sim \log(N)/N$ of the peak emission time remain remarkably robust, even when atomic positions are strongly disordered. In the extreme limit of a homogeneously distributed atomic ensemble, we found that the peak rate is only reduced by a constant factor of approximately $1/4$ and that there is no change in the asymptotic scaling behavior. However, disorder affects the finite-size scaling toward this limit, which explains the difficulty of exploring these effects with most of the previously employed numerical methods that are limited to small atom numbers.

To gain physical understanding of our numerical findings on this robustness, we developed a variational product-state ansatz that yields a physically motivated upper bound on the collective decay rate. The bound $\mathbb{R}_{<}$ in Eq.~\eqref{RstarlowerDef} not only captures the observed superradiant scaling for disorder of varying strength, but also offers conceptual insight into the mechanism behind its robustness. Specifically, it reveals a form of disorder-induced self-organization: instead of synchronizing to align along a global direction as in the weakly disordered case, the atomic spins align coherently along locally determined directions set by the relative propagation phases. This form of spontaneous spin ordering always leads to constructive interference for collective emission, even under strong disorders. Using QSDMF simulations that can provide access to trajectory-level quantities, the spin ordering is confirmed numerically. Interestingly, this spin ordering also reveals the mechanism that underlies the asymmetric photon correlations reported in \cite{CardenasLopezPRL2023}, since the two ordering patterns prefer emission to the left and the right, respectively.

The scalable numerical techniques and the analytic framework, which we have applied and benchmarked here for the study of superradiance in disordered 1D arrays, can be readily generalized for the investigation of collective emission effects in many other settings. Of particular interest in this context is the DTWA approach, which scales efficiently up to large atom numbers and can be adapted for non-Markovian scenarios with minimal additional computational effort. In turn, the QSDMF method provides additional intuitive insights into the actual ordering of atomic dipoles during decay, which can be used in conjunction with the analytically established upper bound $\mathbb{R}_{<}$ as a transparent and computationally inexpensive rule of thumb for superradiance in more general settings. 

\subsection{Relation to Previous Work}

Since Dicke's pioneering work~\cite{DickePhysRev1954}, superradiance has been studied extensively in atomic gases and other extended media~\cite{GrossPhysRep1982,SkribanowitzPRL1973,HaakePRL1979,HaakePRA1979,SchuurmansOC1980,MattarPRL1981,DrummondPRA1982,MacGillivrayPRA1976}. In the continuum limit, where a dense ensemble can be approximated as a continuous medium rather than as discrete radiating dipoles, collective bursts are well established. However, much less is understood about superradiance beyond this limit in strongly disordered discrete systems, where permutation symmetry is absent and the many-body dynamics rapidly becomes intractable at large $N$. While recent studies~\cite{MassonNatComm2022,CardenasLopezPRL2023,MokArxiv2024} have provided interesting arguments about collective decay in extended or disordered arrays based on photon correlations or upper bounds of maximal decay rates, there is no direct large-$N$ characterization, let alone a microscopic mechanism for its robustness.

Building on the recent observation that local observables in superradiant decay can admit an accurate separable description~\cite{ZhangPRL2025}, we address this problem using two complementary large-scale simulation methods together with an analytical approach. Beyond establishing the robustness of the superradiant $N^2$ scaling in the disordered waveguide setting, our central conceptual advance is to identify the mechanism that makes this robustness possible: a spontaneous disorder-aligned spin ordering generated by the competition between the left- and right-propagating decay channels. Relative to previous work, the present results contribute in two main ways.

(i) We provide direct large-$N$ evidence that the characteristic superradiant $N^2$ scaling remains intact even under strong spatial disorder in a bidirectional waveguide. At the same time, the trajectory-based simulations reveal fluctuation-induced effects that are absent in continuum or MF descriptions~\cite{GrossPhysRep1982,SkribanowitzPRL1973,HaakePRL1979,HaakePRA1979,SchuurmansOC1980,MattarPRL1981,DrummondPRA1982,MacGillivrayPRA1976}, most notably the spontaneous selection of one of two ordered spin configurations and the resulting anti-correlation between left- and right-going emission channels. These results show that fluctuations are not merely quantitative corrections here, but are essential for capturing the microscopic physics of the collective burst.

(ii) To our knowledge, the spontaneous spin ordering identified here is a new self-organization mechanism for superradiant decay in waveguide QED, especially in the strongly disordered regime considered here. For zero or weak spatial disorder, phase synchronization into a nearly uniform configuration is well known in spontaneous emission~\cite{GrossPhysRep1982} and under incoherent pumping~\cite{ZhuNJP2015}. Likewise, for purely chiral emission with a single decay channel, the propagation phases can be absorbed into a redefinition of spin operators and therefore do not affect the dynamics (see e.g.~\cite{BonifacioPRA1971b,GrossPhysRep1982,LiedlPRX2024}). In contrast, in the bidirectional case studied here, the competition between two decay channels gives rise to two emergent spin configurations whose phases adapt to the random atomic locations and thereby preserve constructive interference. We further show that related spin ordering also appears in regularly spaced arrays, suggesting that this mechanism is not restricted to disordered ensembles and may arise more broadly in cooperative light-matter dynamics.

\subsection{Outlook}

Looking forward, several promising directions emerge. While our current model involves two decay channels arising from the bidirectional waveguide, extensions to systems with multiple decay channels may reveal new forms of collective phenomena. Though the synchronization process can be captured quantitatively by our semiclassical EOMs, a transparent and qualitative theory describing the dynamical formation of spin ordering would also be interesting for future research. Additionally, studying superradiance in engineered environments---such as squeezed reservoirs, structured bath, or chiral waveguides---may uncover richer phenomena of coherence and self-organization. The analytical and numerical frameworks introduced here are broadly adaptable for future studies of disorder, coherence, and collectivity in many-body quantum optics. Beyond its fundamental interest, the predicted robustness of these collective quantum effects against various forms of local disorders is also of significant practical importance for potential applications that rely on collective radiation effects. This includes the efficient generation of metrologically useful states~\cite{WangPRL2014,PaulischPRA2019,PerarnauLlobetQST2020,Arya2024,BelliardoArxiv2026}, the development of ultra-narrow superradiant lasers~\cite{MeiserPRL2009,BohnetNature2012,NorciaSciAdv2016}, as well as recent proposals for superradiant quantum batteries \cite{CampaioliRMP2024}. Our findings suggest that such applications exhibit significantly greater resilience to imperfections than commonly assumed. More broadly, the existence of cooperative effects that remain robust even under strong disorder, such as the spin ordering identified here, challenges the conventional intuition that imperfections destroy nontrivial collective quantum behavior in realistic platforms.


\section*{Data availability}

Numerical data for the figures are available at~\cite{Zhang_data_2025}.

\begin{acknowledgments}
We thank Constanze Bach, Iman Marvian, Arno Rauschenbeutel, and Philipp Schneeweiss for helpful discussions. We acknowledge support by the Deutsche Forschungsgemeinschaft (German Research Foundation)–522216022 and by the ERC project ‘ConsQuanDyn’ (No.\,851161). D.\,M.\,acknowledges support from the Novo Nordisk Foundation under Grants No.\,NNF22OC0071934 and No.\,NNF20OC0059939. This research is part of the Munich Quantum Valley, which is supported by the Bavarian state government with funds from the Hightech Agenda Bayern Plus.
\end{acknowledgments}


\appendix


\section{Finite-Size Scaling of the Decay Rates}\label{SecSingleExcitationUpperBound}

The total collective decay rate $\mathcal{R}$ introduced in Eq.~\eqref{EqR} can in general be written as 
\begin{equation}\label{Rmatrix}
    \mathcal{R} = \sum_{ij} \hat{R}_{ij} \braket{\sigma_{i}^{+} \sigma_{j}^{-}} = \sum_{k} \Gamma_{k} \braket{\Omega_{k}^{\dagger} \Omega_{k}},
\end{equation}
where $\hat{R}$ is a Hermitian matrix defined by the jump operators in Eq.~\eqref{JumpOp}, $\Gamma_k$ are its eigenvalues, and $\Omega_{k} = \sum_{j} U_{kj} \sigma_{j}^{-}$ are the corresponding decay modes after diagonalization of $\hat{R}$ with a unitary transformation $U$. While the eigenvalues are only directly related to the decay rates of the first few emissions, an upper bound of the maximal decay rate can be written out as (see Appendix~\ref{AppdxOtherBounds})
\begin{equation}\label{EqP}
\mathcal{R}_{\star} \leq   \frac{3}{2} N \Gamma_{+}.
\end{equation}
Therefore, the upper bound indicates that $\mathcal{R}_\star\sim N \Gamma_{+}$, which means the finite-size scaling of $\mathcal{R}_\star$ could be hinted in the finite-size scaling of $\Gamma_{+}$. Though it is a priori unclear if $\mathcal{R}_\star\sim N\Gamma_{+}$ holds, our numerical results confirm so.

In the current case, $\hat{R}$ in Eq.~\eqref{Rmatrix} can be written as
\begin{equation}\label{RMatrix}
    \hat{R} = \frac{\gamma}{2}N ( \ket{\psi_{1}} \bra{\psi_{1}} + \ket{\psi_{2}} \bra{\psi_{2}} ),
\end{equation}
with $\ket{\psi_{1}} = (e^{i \xi_{1}}, e^{i\xi_{2}}, \ldots, e^{i \xi_{N}})^{T}/\sqrt{N}$ and $\ket{\psi_{2}} = (e^{-i\xi_{1}}, e^{-i\xi_{2}}, \ldots, e^{-i\xi_{N}})^{T}/\sqrt{N}$, which originate from the coupling of the atoms to the right- and left-going channels, respectively. 
%
These two states are not orthogonal and we can define
\begin{equation}
    \ket{\psi_{2}} = c \ket{\psi_{1}} + \sqrt{1-|c|^2} \ket{\psi_{1}^{\perp}},  
\end{equation}
where $\ket{\psi_{1}^{\perp}}$ is a state orthogonal to $\ket{\psi_{1}}$.
The overlap $c$ is the sum of the random numbers $e^{-i2\xi}$ [cf.~Eq.~\eqref{cDef}], which at large $N$ obeys a Gaussian distribution according to the central limit theorem. In the basis $\{\ket{\psi_1},\ket{\psi_1^\bot}\}$,  $\hat{R}$ has the form
\begin{equation}\frac{\gamma}{2}N 
  \begin{pmatrix}
1+|c|^2 & c \sqrt{1-|c|^2}  \\
c^{*} \sqrt{1-|c|^2} & 1-|c|^2 
\end{pmatrix},
\end{equation}
of which the diagonalization gives two eigenvalues $ \Gamma_{\pm} = \frac{\gamma}{2}N ( 1 \pm |c|)$. Then our original master equation \eqref{ME0} can be rewritten as Eq.~\eqref{TwoChannelLindbladian} with two orthogonal modes.

Since $\xi_{j}$ is uniformly distributed in $[-\Theta/2, \Theta/2)$, we know $ \Re e^{-i2\xi}$ and $ \Im e^{-i2\xi} $  have the means and variances shown in Eq.~\eqref{MeanVar}. Thus, for large $N$, $c$ obeys a Gaussian distribution with $\Re c \sim \mathcal{N}(\mu_{a}, \sigma_{a}^2/N)$ and $\Im c \sim \mathcal{N}(\mu_{b}, \sigma_{b}^2/N)$. In the thermodynamic limit $N\rightarrow\infty$, we then obtain Eq.~\eqref{NSScaling}.

To evaluate the finite-size scaling, we consider large but finite $N$. To get an analytical understanding, we can write $\Re c = \mu_{a} + \delta_{a}$ and $\Im c =\delta_{b}$, with $\delta_{a/b}$ being small Gaussian fluctuations $\sim N^{-1/2}$. Then we see that 
\begin{equation}\label{c1Expansion}
    |c| = \sqrt{\mu_{a}^2 + 2\mu_{a}\delta_{a} + (\delta_{a}^2+\delta_{b}^2)}.
\end{equation}
For the values of $\Theta=n\pi$, where $\mu_a=0$, we find $|c| = \sqrt{(\delta_{a}^2+\delta_{b}^2)} \sim |\delta_{a/b}| \sim N^{-1/2} $. This explains the scaling $E[\Gamma_{+}] = \frac{1}{2} \gamma N \left( 1 + \mathcal{O}\left(N^{-1/2}\right) \right)$ for the finite-size corrections, in accordance with the exact results in Eq.~\eqref{NSScalingThetaNPi}. For any other $\Theta \neq n\pi$, where $\mu_{a} \neq 0$, we can make the expansion
\begin{equation} \label{c1ExpansionNotNPi}
    |c| = |\mu_{a}| \left(1 + \frac{\delta_{a}}{\mu_{a}} +\frac{1}{2} \frac{\delta_{b}^2}{\mu_{a}^2}  + \ldots  \right)
\end{equation}
to obtain
\begin{equation}
    E[|c|] =  |\mu_{a}|   + \frac{1}{2|\mu_{a}|} E[\delta_{b}^2] \sim  |\mu_{a}|   + \mathcal{O} (N^{-1}).
\end{equation}
Therefore, in this case we find a finite-size correction of the form 
\begin{equation}
    E[\Gamma_{+}] = \frac{1}{2} \gamma N \left( 1 + |\mu_{a}|   + \mathcal{O} \left(N^{-1}\right) \right),
\end{equation}
consistent with Eq.~\eqref{NSScaling}.

Therefore, the convergence is slower at special points $\Theta = n\pi$, where $\Gamma_{+} = \Gamma_{-}$. This originates mathematically from the fact that the overlap $c$ approaches $0$ for $\Theta=n\pi$. We note that, though the bound $\mathbb{R}_{<}$ in Sec.~\ref{secProductAnsatz} is much tighter, it always has a finite-size correction $\sim \mathcal{O}(N^{-1})$. The reason could be that the product state ansatz is valid for large $N$, which therefore fails to capture the intermediate-size crossover. A similar difference in the finite-size scaling is numerically observed for the actual peak decay rate $\mathcal{R}_\star$, indicating the validity of the relation $\mathcal{R}_\star\sim N \Gamma_+$.




\section{Markovian Limit of DTWA} \label{DTWAappendix}
In Sec.~\ref{subsecDTWA}, we have presented the stochastic EOM for a set of TLAs coupled to two damped cavity modes. In numerical simulations, we can then choose a sufficiently large damping rate $\kappa$ to recover an effective Markovian dynamics for the TLAs only. When only Markovian dynamics is concerned, an adiabatic elimination of the cavity modes can be carried out analytically, as discussed in the following. From a numerical perspective, eliminating the cavity modes avoids a costly simulation of the bosonic amplitudes on a very fast timescale and enforces the Markovian limit assumed in the original master equation. However, keeping the bosonic modes as dynamical degrees of freedom has the advantage that it can be readily extended to a non-Markovian setting, as discussed in Sec.~\ref{secNonMarkov}. 

We start by introducing a time increment $\Delta t$, which is short with respect to the dynamics of the TLAs, but long compared to $\kappa^{-1}$. By integrating the EOM for the spin variables in Eq.~\eqref{SpinEOM} over this time interval, we obtain, for example, 
\begin{equation}
\begin{split}
        \Delta s_{j}^{-}(t)& \simeq i \frac{\gamma}{2}\sum_{l\neq j} \sin(|\xi_j-\xi_l|) s_j^z s_l^- \Delta t\\
        &+ \frac{g}{\sqrt{2}} s_{j}^{z} \left(\Delta \alpha_{\text{R}}(t) e^{i\xi_j}  +  \Delta \alpha_{\text{L}}(t) e^{-i\xi_j} \right),
\end{split}
\end{equation}
where $\Delta \alpha_{\text{R,L}}(t)=\int_t^{t+\Delta t} \alpha_{\text{R,L}}(s)\,ds$ are the corresponding increments of the cavity amplitudes. From the formal solutions of Eq.~\eqref{AlphaSDE}, we have 
\begin{equation}\label{AlplaAdiabatic}
\begin{split}
\alpha_{\text{R,L}}(t) =&\frac{g}{\sqrt{2}}  \int_{0}^{t} ds \, e^{-\frac{\kappa}{2}(t-s)} \tilde J_{\rm R,L}(s) \\&+ \sqrt{\frac{\kappa}{2}} \int_{0}^{t} e^{-\frac{\kappa}{2}(t-s)} dW_{\rm R,L}(s)\\
        \approx & \frac{\sqrt{2}g}{\kappa} \tilde J_{\rm R,L}(t)+ \sqrt{\frac{2}{\kappa}}\Xi_{\rm R,L}(t),
\end{split}
\end{equation}
where the last approximation is valid when $\kappa/(\gamma N) \gg 1$. Here, we have introduced the noise process
\begin{equation}
\Xi_{\rm R,L}(t)=\frac{\kappa}{2} \int_{0}^{t} e^{-\frac{\kappa}{2}(t-s)} dW_{\rm R,L}(s),
\end{equation}
which is $\delta$-correlated on timescales longer than $\kappa^{-1}$. From Eq.~\eqref{AlplaAdiabatic}, we see that the occupation number in the cavity is of order $\mathcal{R}/\kappa$, which is $\sim \mathcal{O}(1)$ and $\sim \mathcal{O}(N)$ for $\mathcal{R} \sim \gamma N$ initially and $\mathcal{R} \sim \gamma N^2$ around superradiance respectively, when $\kappa \sim \gamma N$.  By defining the corresponding coarse-grained increments, $\Delta W_{\rm R,L}(t)=\int_t^{t+\Delta t}\Xi_{\rm R,L}(s)$, we obtain the effective stochastic updates for the spin variables, 
\begin{equation}
\begin{split}
        \Delta s_{j}^{-}(t) \simeq & \frac{\gamma}{2} \sum_{l} s_{j}^{z} s_{l}^{-} e^{i|\xi_{j} - \xi_{l}|}  \Delta t\\
        & +\sqrt{\frac{\gamma}{4}}s_{j}^{z}  \left(\Delta W_{\rm R}(t)e^{i\xi_j}  +  \Delta W_{\rm L}(t) e^{-i\xi_j} \right),
\end{split}
\end{equation}
together with a similar equation for the $\Delta s_j^z(t)$. On the timescale of the spin variables, we can now treat all increments as infinitesimal to recover a regular set of stochastic differential equations. These are, however, of Stratonovich type and must be converted into Ito form~\cite{GardinerBookStochastic} for numerical simulations. We finally end up with 
\begin{equation}\label{SpinEOMAfterElimination}
\begin{split}
    d s_{j}^{-} =& - \frac{\gamma}{2} s_{j}^{-} dt + \frac{\gamma}{2} \sum_{l} s_{j}^{z} s_{l}^{-} e^{i|\xi_{j} - \xi_{l}|} dt + \sqrt{\frac{\gamma}{2}} s_{j}^{z} dW_{j}   \\
        d s_{j}^{z} =& - \gamma s_{j}^{z} dt - \gamma \sum_{l} \left( s_{j}^{+}s_{l}^{-} e^{i|\xi_{j} - \xi_{l}|} + \text{c.c.} \right) dt   \\
        & - \sqrt{2\gamma} \left( s_{j}^{+} dW_{j} + \text{c.c.} \right)  ,
\end{split}
\end{equation}
where the $dW_{j} = (e^{i\xi_{j}} dW_{\text{R}} + e^{-i\xi_{j}} dW_{\text{L}})/\sqrt{2} $ are a set of $N$ \emph{interdependent} complex Wiener increments, defined in terms of the two independent increments $dW_{\text{R}}$ and $dW_{\text{L}}$. 

Note that, in Eq.~\eqref{SpinEOMAfterElimination}, with the expression for $\alpha_{\text{{R/L}}}$ being semiclassical, there is no contraction for $l=j$. For $N=1$, the EOM is \emph{not} the one we expect from MF theory, since it ignores the obvious contraction of the form $\sigma_{j}^{\alpha} \sigma_{j}^{\beta}$ at the same site. But spin length is conserved in this way. This on-site error becomes negligible as long as there are enough spin correlations, which is the case around superradiance.


\section{Numerical Benchmarking of Different Methods}\label{CompareMethods}

\begin{figure}[t]
	\center
\includegraphics[width=\columnwidth]{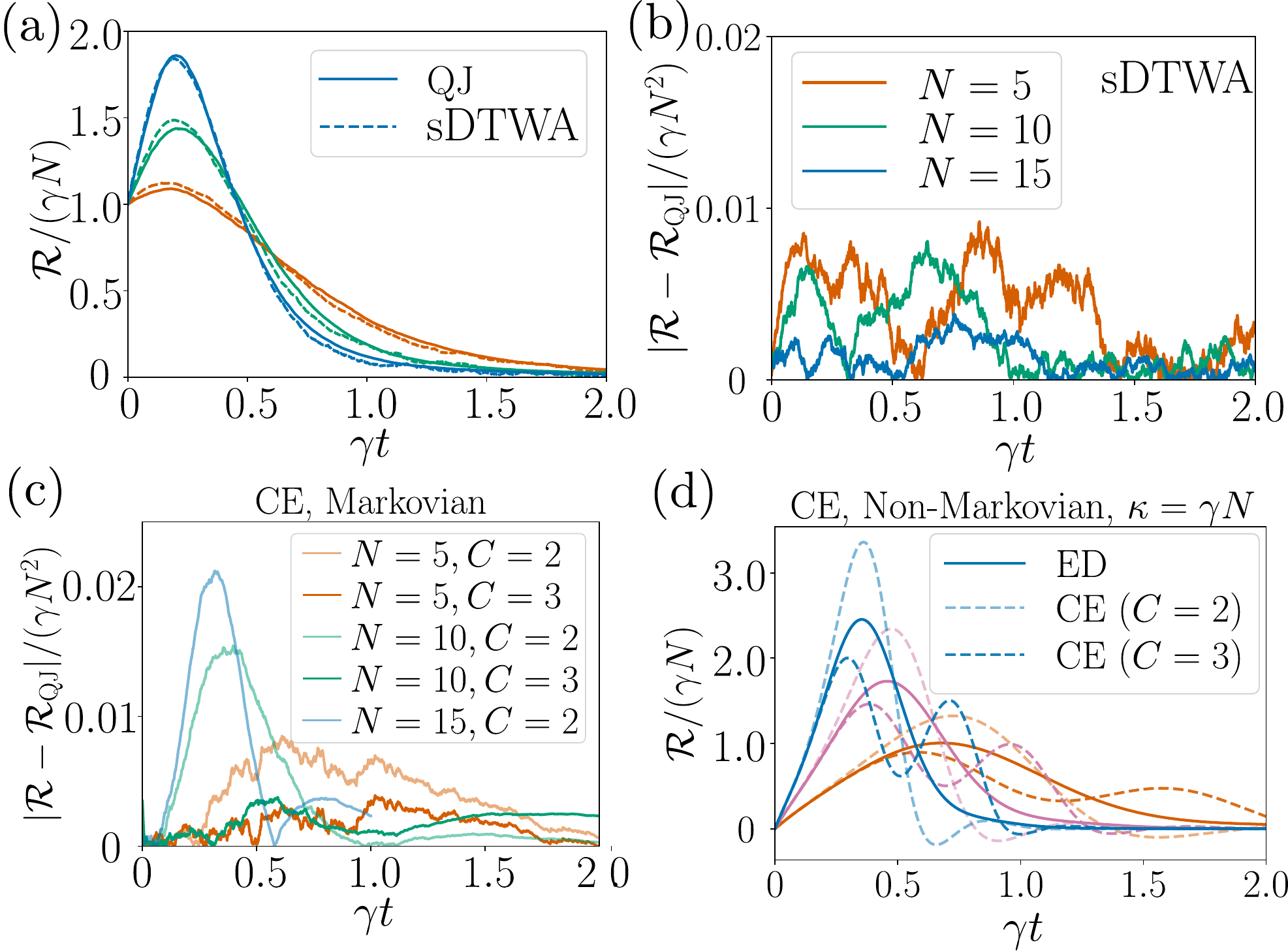}
	\caption{Comparison of different numerical methods. Error in computing $\mathcal{R}/(\gamma N^2)$ using (a) sDTWA, (b) QSDMF, and (c) CE (with order $C=2$ or $3$), benchmarked against QJ simulations, for system sizes $N=5, 10, 15$. All averages have been taken over $\mathcal{N}=10^3$ trajectories. (d)  
    Comparison between ED and CE for simulating $N=5,10,15$ atoms coupled with one lossy cavity with $\kappa=\gamma N$.}
	\label{FigErrorComparison}
\end{figure}

Here, in complement to Sec.~\ref{secMethodsCompareAnalysis}, we show numerical benchmarks of three approximate numerical methods -- a variant of DTWA in spherical coordinates (sDTWA), quantum state diffusion with product state approximation (QSDMF), and cumulant expansion (CE) -- against exact simulations using either quantum jump (QJ) or exact diagonalization (ED).  In the Markovian case described by Eq.~\eqref{ME0}, as shown in Fig.~\ref{FigErrorComparison} (a-c), all these methods are applicable and accurate for computing the superradiant peaks. 

For the non-Markovian case, we consider $N$ TLAs that are coupled uniformly to one lossy cavity with decay rate $\kappa = \gamma N$, such that the permutation symmetry permits exact simulations with moderate atom numbers of $N=5,10,15$. Since the cavity can be truncated to keep the first $(N+1)$ levels, the Hilbert space dimension needed for exact diagonalization (ED) is $(N+1)^2$. In this setting, we compute the collective decay rate as $\mathcal{R} (t) = - \frac{d}{dt} \sum_{j} \braket{\sigma_{j}^{z}}/2$. As can be seen in Fig.~\ref{FigNMDecay}(a), even when $\kappa = \gamma N$, results from DTWA agree with those from ED remarkably well. In contrast, CE with order $C=3$ still shows significantly larger errors [see Fig.~\ref{FigErrorComparison}(d)]. We thus conclude that our DTWA formalism is the only one of these four methods that can deal with non-Markovian effects, while remaining efficient and accurate.

\section{Vanishing Mutual Information in Disordered Waveguide QED}\label{AppdxMutualInfo}

To further substantiate the validity of QSDMF in the large-$N$ limit, we verify the finding of Ref.~\cite{ZhangPRL2025}---namely that the mutual information along QSD trajectories vanishes with increasing system size---also in the present, strongly disordered setting. This property justifies the validity of the product-state approximation at the trajectory level.

\begin{figure}[t]
	\center
\includegraphics[width=\columnwidth]{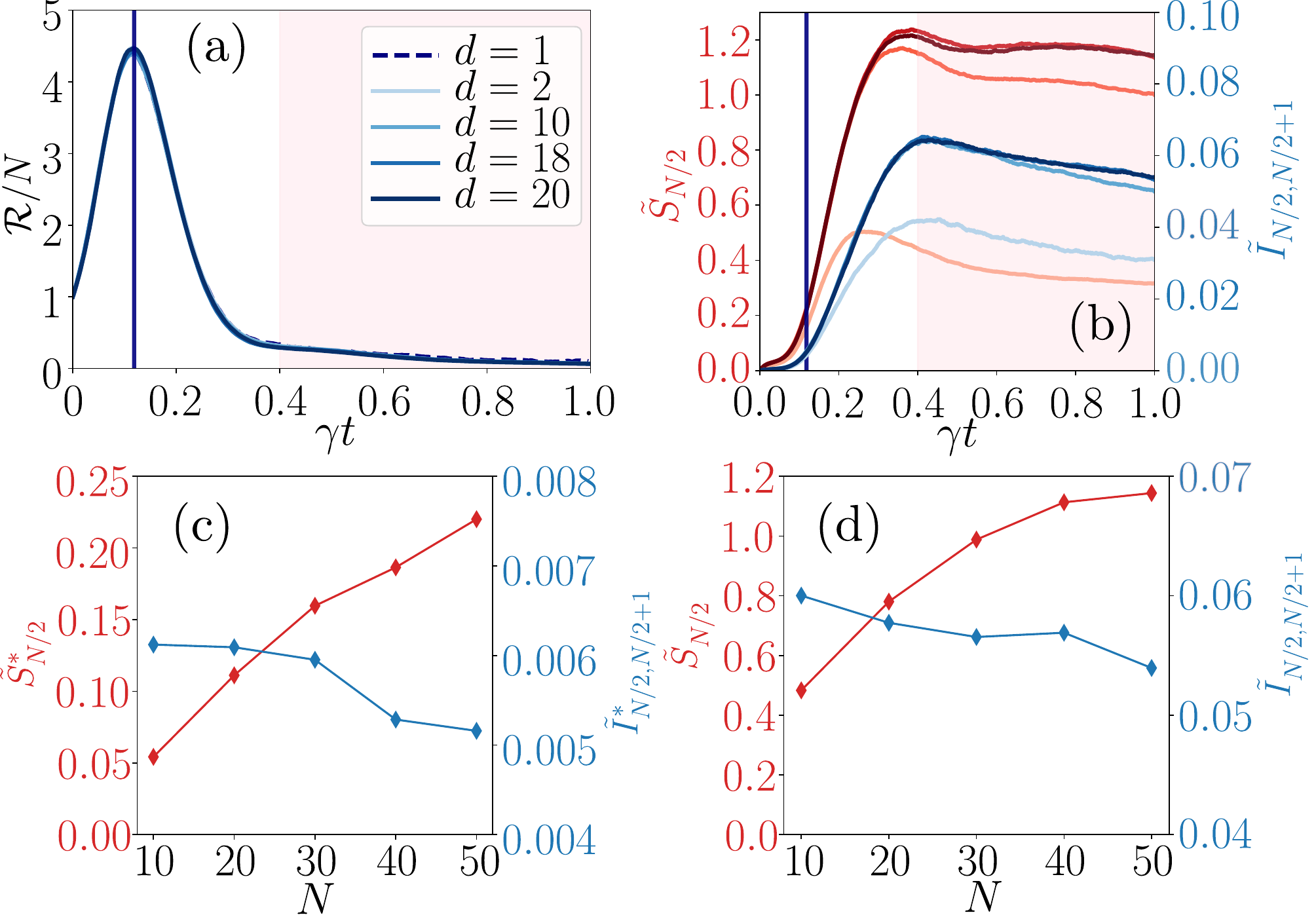}
	\caption{Vanishing mutual information. MPS simulations are performed for QSD trajectories with increasing bond dimension $d$ at $\Theta=2\pi$. Converged results are already obtained for a bond dimension of $d=20$. (a) Time evolution of the decay rate $\mathcal{R}/N$ for $N=50$. The vertical line marks the superradiant peak time $t_{\star}$, and the pink shaded region denotes the subradiant regime. The difference between the QSDMF result ($d=1$) and the numerically exact result is negligible, thereby verifying the accuracy of QSDMF for $N=50$. (b) Time evolution of the trajectory-averaged bipartite entanglement entropy $\tilde{S}_{N/2}$ and the trajectory-averaged mutual information between the two central TLAs, $\tilde{I}_{N/2,N/2+1}$, for $N=50$. (c) and (d) Scaling of $\tilde{S}_{N/2}$ and $\tilde{I}_{N/2,N/2+1}$ with $N$ at the superradiant peak time $t_{\star}$ and at $\gamma t=1$ in the subradiant regime, respectively. In both cases, $d=20$ is used, and the mutual information between the two TLAs decreases as $N$ increases. All simulations have been performed with a time step $\gamma\delta t=2\times10^{-4}$ and using $\mathcal{N}=10^3$ trajectories. }
	\label{FigQSDMPS}
\end{figure}

In Fig.~\ref{FigQSDMPS}, we perform MPS simulations of the QSD trajectories for the representative case $\Theta=2\pi$. For $N=50$, the trajectory-averaged bipartite entanglement entropy $\tilde{S}_{N/2}$ requires a bond dimension $d\sim 20$ for convergence. In contrast, local observables, such as the decay rate $\mathcal{R}$, are already reproduced by the MF approximation with negligible error. The reason lies in the fact that the trajectory-averaged mutual information between a pair of spins $\tilde{I}_{j,j^{\prime}}$ becomes vanishingly small with increasing $N$. In Fig.~\ref{FigQSDMPS}(c) and (d), we show $\tilde{I}_{j,j^{\prime}}$ for the central pair of TLAs at the superradiant peak time $t=t_{\star}$ and at $t=1/\gamma$ in the subradiant regime, respectively. In both cases, although $\tilde{S}_{N/2}$ grows with $N$, $\tilde{I}_{N/2,N/2+1}$ remains far below unity and continues to decrease. Other spin pairs behave similarly. These results, consistent with Ref.~\cite{ZhangPRL2025}, show that QSDMF remains highly accurate for local observables at large $N$.

\section{Gaussian-Distributed Positions}\label{AppdxGaussian}

\begin{figure}[t]
	\center
\includegraphics[width=\columnwidth]{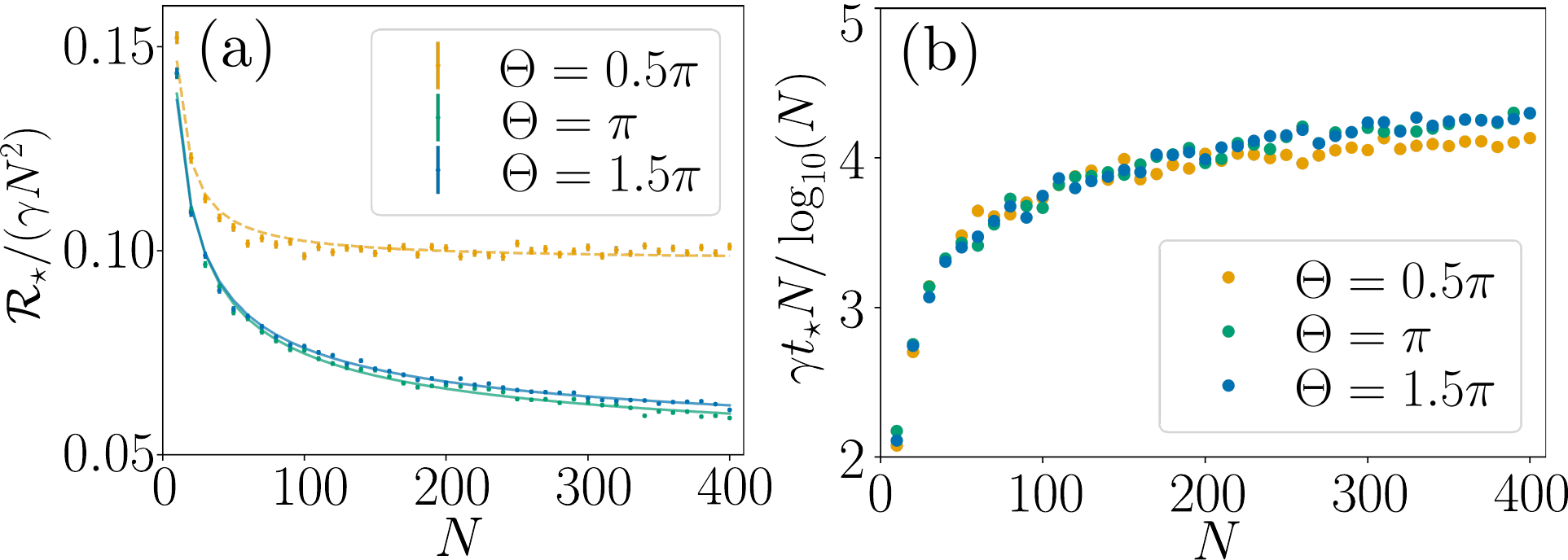}
	\caption{Superradiance with Gaussian-distributed positions. Scaling of (a) the superradiant rate $\mathcal{R}/(\gamma N^2)$ and (b) the burst time $t_{\star}\gamma N/\log(N)$, for Gaussian-distributed positions with disorder strengths $\Theta = \pi/2, \pi, 3\pi/2$. The plots demonstrate a convergence to an $N^2$-scaling, but with different finite-size corrections. We find that the data can be fitted with the function $N^2 ( r_0 + r_1 N^{-1} )$ for $\Theta=\pi/2$ (dashed line), and $N^2 ( r_0 + r_1^{\prime} N^{-1/2} )$ for $\Theta=\pi$ and $\Theta = 3\pi/2$ (solid lines).    Results are averaged over $\mathcal{N}=10^3$ trajectories.  }
	\label{FigGaussian}
\end{figure}

For completeness, we here consider the case of Gaussian position fluctuations instead of the uniform distribution considered in the main text.
Specifically, let the propagation phases $\{\xi_j\}$ obey a Gaussian distribution $\mathcal N(0, \Theta^2/4)$, with standard deviation $\Theta/2$ for comparison with the uniform case considered above. Then, the quantities $ \Re e^{-i2\xi}$ and $ \Im e^{-i2\xi} $ have the following mean values and variances:
\begin{align}
\tilde{\mu}_{a} & = e^{-\Theta^2/2},  & \tilde{\sigma}_{a}^{2} &= \frac{1}{2}\left(1 - 2 e^{-\Theta^2} + e^{-2\Theta^2}\right) , \nonumber \\
\tilde{\mu}_{b} & =0, &  \tilde{\sigma}_{b}^{2} &=  \frac{1}{2}\left(1 - e^{-2\Theta^2}\right).  \label{GaussianMeanVar}
\end{align}
Consequently, in the large $N$ limit, $c$ obeys a Gaussian distribution with $\Re c \sim N(\tilde{\mu}_{a}, \tilde{\sigma}_{a}^2/N)$ and  $\Im c \sim N(\tilde{\mu}_{b}, \tilde{\sigma}_{b}^2/N)$. Following the same procedure as in Appendix~\ref{SecSingleExcitationUpperBound}, we obtain for  $N\rightarrow\infty$, 
\begin{equation} \label{NSScalingGaussian}
  \Gamma_{\pm} \rightarrow \frac{1}{2} \gamma N ( 1 \pm \tilde{\mu}_{a}) = \frac{1}{2} \gamma N \left( 1 \pm e^{-\Theta^2/2} \right).
\end{equation}
For small $\Theta$, we have $\tilde{\mu}_a \sim  1$, and using the expansion in Eq.~\eqref{c1ExpansionNotNPi}, we obtain the scaling 
\begin{equation}\label{EqGaussianSmallDisorder}
    E[\Gamma_{+}] = \frac{1}{2} \gamma N \big( 1 + \tilde{\mu}_{a}   + O (N^{-1}) \big),
\end{equation}
with a finite-size correction of order $1/N$ for $\Gamma_{+}/N$. For large $\Theta$ and finite $N$, i.e., $\Theta \gg \sqrt{\log(N)}$, we encounter the relation $0<\tilde{\mu}_a \ll \delta_{a}$, and, according to Eq.~\eqref{c1Expansion},
\begin{equation}
    E[|c|] \approx E[\sqrt{(\delta_{a}^2+\delta_{b}^2)}] \sim E[|\delta_{a/b}|] \sim N^{-1/2}.
\end{equation}
Therefore, under this condition we find a slow finite-size scaling 
\begin{equation}\label{EqGaussianLargeDisorder}
    E[\Gamma_{+}] \approx \frac{1}{2} \gamma N \left( 1    + O (N^{-1/2}) \right).
\end{equation}

In Fig.~\ref{FigGaussian}, we plot $\mathcal{R}_{\star}/(\gamma N^2)$ as a function of $N$ for Gaussian distributions with $\Theta=\pi/2, \pi, 3\pi/2$. For $\Theta=\pi/2$, $\tilde{\mu}_{a} \approx 0.29$, Eq.~\eqref{EqGaussianSmallDisorder} predicts a finite-size scaling $\sim O(N^{-1})$, which agrees with our numerical data. As predicted in Eq.~\eqref{EqGaussianLargeDisorder}, $\Theta=\pi$ and $\Theta=3\pi/2$ give $\tilde{\mu}_{a} \approx 7\times 10^{-3}$ and $\tilde{\mu}_{a} \approx 1.5 \times 10^{-5}$, respectively, which leads to a finite-size scaling $\sim O(N^{-1/2})$ as observed in numerical simulations with $N\leq 400$. 

Following Sec.~\ref{secProductAnsatz}, we can derive the upper bound
\begin{equation}
    \mathbb{R}_{<} = \frac{\gamma N^2}{8}\left(1 + e^{-\Theta^2}\right),
\end{equation}
for a Gaussian distribution, which tightly bounds the numerical data from above. Similar to the case with homogeneous distribution, $\mathbb{R}_{<}$ always has finite-size correction $\sim N^{-1}$, but does not capture the transient $N^{-1/2}$ scaling.

\section{A Loose Upper Bound}
\label{AppdxOtherBounds}

For a traceless $2$-local spin-$1/2$ Hamiltonian $H$ with negative ground state energy, Lieb \cite{lieb_classical_1973} proved that the ground-state energy can be bounded from below with $9\braket{H}_{\text{prod,min}}$, where $\braket{H}_{\text{prod,min}}$ is the lowest energy of $H$ obtained by minimizing over all product states. Then Bravyi {\it et al} \cite{bravyi_approximation_2019} further proved that ground state energy can be lower bounded by $6 \times \min \{ \braket{H}_{\text{prod,min}}, \braket{-H}_{\text{prod,min}} \}$. Recently, these bounds were applied to the problem of superradiance \cite{MokArxiv2024} on bounding the largest eigenvalue of $\hat{\mathcal{R}}$. Based on the finding in \cite{ZhangPRL2025}, we argue that the bound $\mathbb{R}$ using product states is sufficient to upper bound the superradiant rate. Here, following \cite{MokArxiv2024}, we show the upper bound of the largest eigenvalue of $\hat{\mathcal{R}}$ for completeness, and to show that it far overestimates the actual decay rate. 

Here, the operator for the total decay rate has the form
\begin{equation} 
    \hat{\mathcal{R}} = \sum_{i} R_{ii} \sigma_{i}^{+} \sigma_{i}^{-} +  \sum_{i,j\neq i} R_{ij} \sigma_{i}^{+} \sigma_{j}^{-} \equiv \hat{\mathcal{R}}_1 + \hat{\mathcal{R}}_2,
\end{equation}
where $\hat{\mathcal{R}}_2$ is a traceless $2$-local operator.
In our case, 
\begin{equation}
  R_{ij} = \gamma \cos( \xi_{i} - \xi_{j} ).
\end{equation} 
Then we know the largest eigenvalue of $\hat{\mathcal{R}}$ is lower bounded by
\begin{equation}
   \lambda_{\max}(\hat{\mathcal{R}}) \geq  \mathbb{R},
\end{equation}
where $\mathbb{R}$ is bound from our product state ansatz in Sec.~\ref{SecUpperBoundR}.

As for the upper bound of $\lambda_{\max}(\hat{\mathcal{R}})$, since $\braket{\sigma_{i}^{+}\sigma_{i}^{-}} \leq 1$ , we know
\begin{equation}
    \lambda_{\max}(\hat{\mathcal{R}}) \leq \gamma N + \lambda_{\max}(\hat{\mathcal{R}}_2). 
\end{equation}
Using the theorem in \cite{lieb_classical_1973}, we know
\begin{equation}
  \lambda_{\max}(\hat{\mathcal{R}}_2) \leq 9 \lambda_{\text{prod}, \max}(\hat{\mathcal{R}}_2),
\end{equation}
where $\lambda_{\text{prod}, \max}(\hat{\mathcal{R}}_2)$ denotes the maximum value $\braket{\hat{\mathcal{R}}_2}$ over all product states.

To bound $\lambda_{\text{prod}, \max}(\hat{\mathcal{R}}_2)$ from above,  we can write
\begin{equation}
\begin{split}
    \lambda_{\text{prod}, \max}(\hat{\mathcal{R}}_2) &=  \sum_{i,j\neq i} \gamma \cos( \xi_{i} - \xi_{j} ) \braket{\sigma_{i}^{+} \sigma_{j}^{-}}_{\text{prod}} \\
    &=  \sum_{i,j\neq i} \frac{\gamma}{4}  \cos( \xi_{i} - \xi_{j} ) ( s_{i}^{x} s_{j}^{x} + s_{i}^{y} s_{j}^{y} ) ,
\end{split}
\end{equation}
with a constraint $(s_{i}^{x})^2 + (s_{i}^{y})^2 \leq 1$.
As in Ref.\,\cite{MokArxiv2024}, we can introduce $\hat{S}$ with $S_{ij} = ( s_{i}^{x} s_{j}^{x} + s_{i}^{y} s_{j}^{y} )$, which leads to
\begin{equation}\label{lambdaprodmaxR}
\begin{split}
    \lambda_{\text{prod}, \max}(\hat{\mathcal{R}}_2)  &=\frac{1}{4} \sum_{i,j\neq i} R_{ij} S_{ji} = \frac{1}{4}\Tr\left[\left(\hat{R} - \gamma I\right) \hat{S} \right] \\
    &\leq  \frac{1}{4} (\Gamma_{\max} -\gamma) \Tr\left[ \hat{S} \right] \leq \frac{1}{4} (\Gamma_{\max} -\gamma) N 
\end{split}
\end{equation}
where the second last inequality comes from the fact that $\hat{S} \geq 0$.

Since $\hat{\mathcal{R}} \geq 0$, we know $\braket{\hat{\mathcal{R}}} = \braket{\hat{\mathcal{R}}_1} + \braket{\hat{\mathcal{R}}_2} \geq 0$. From $\braket{\hat{\mathcal{R}}_1}\leq \gamma N$, we know $-\braket{\hat{\mathcal{R}}_2} \leq \gamma N$, which bounds maximal eigenvalue of $-\hat{\mathcal{R}}_2$ by  $\gamma N$ from above. That is, we can further have \cite{bravyi_approximation_2019} 
\begin{equation}
\lambda_{\max}(\hat{\mathcal{R}}_2) \leq 6 \lambda_{\text{prod}, \max}(\hat{\mathcal{R}}_2).
\end{equation}
We then have an upper bound of  $\lambda_{\max}(\hat{\mathcal{R}})$ given by
\begin{equation}\label{lambdamaxR}
\lambda_{\max}(\hat{\mathcal{R}}) \leq \gamma N + 6 \lambda_{\text{prod}, \max}(\hat{\mathcal{R}}_2).
\end{equation}
Combining Eq.~\eqref{lambdaprodmaxR} and \eqref{lambdamaxR}, we have 
\begin{equation}\label{RGreater}
\begin{split}
    \lambda_{\max}(\hat{\mathcal{R}}) \leq \frac{3}{2} \Gamma_{\max} N - \frac{1}{2} \gamma N \rightarrow  \frac{3\gamma}{4} N^2 \left( 1 + \left|\frac{\sin\Theta}{\Theta} \right| \right),
\end{split}
\end{equation}
of which the RHS we denote as the upper bound $\mathbb{R}_{>}$, which overestimates the actual decay rate by more than a factor of $6$.

\section{Neglecting the Hamiltonian}\label{IgnoreAtomInteraction}

\begin{figure}[t]
	\center
\includegraphics[width=\columnwidth]{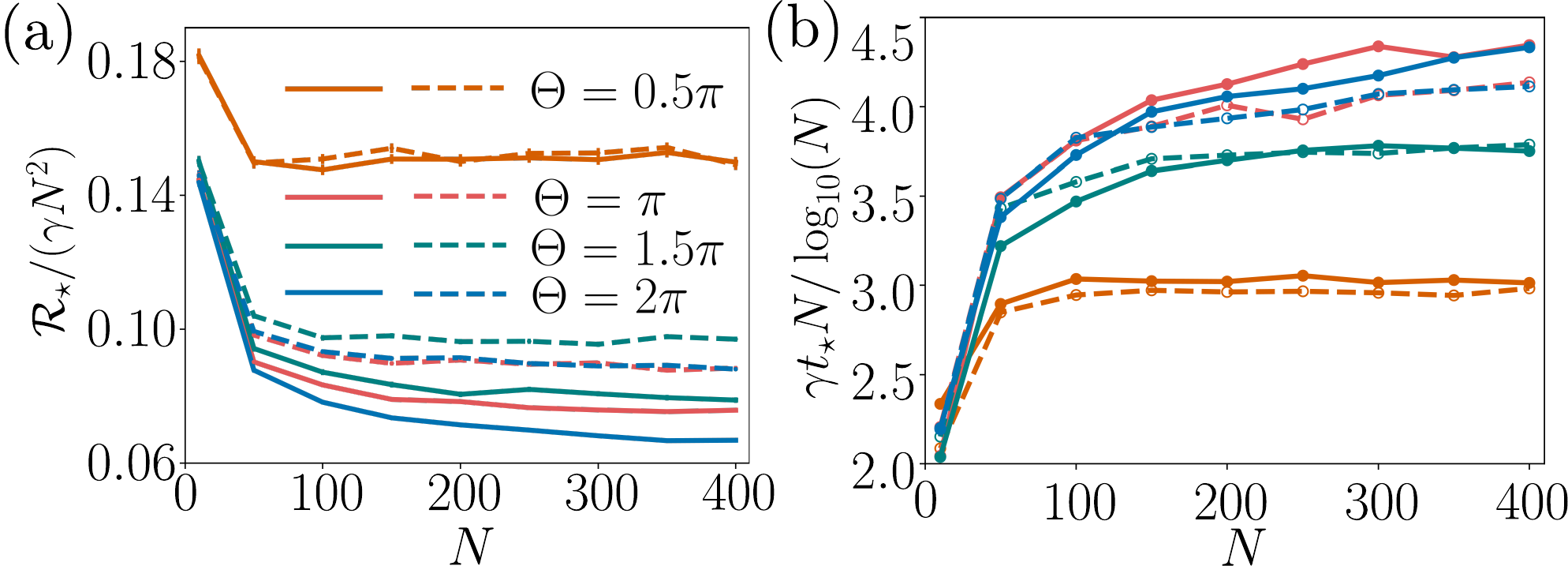}
	\caption{Comparison of superradiant scaling with (solid lines) and without (dashed lines) the coherent Hamiltonian $H$. (a) Plot of the superradiant rate $\mathcal{R}_{\star}/(\gamma N^2)$ and (b) the burst time $t_{\star}$ as a function of $N$ and for different disorder strengths. The Hamiltonian part of the evolution only slightly changes the prefactor of the $N^2$-scaling, but not the scaling itself.}
	\label{FigNoH}
\end{figure}

In Fig.~\ref{FigNoH}, we show results of DTWA simulations without the Hamiltonian $H$. Since the jump operators are the same as those of the original problem, the product state ansatz thus predicts the same upper bound $\mathbb{R}$. Indeed, it can be seen that removing $H$ only increases the prefactor of the $N^2$-scaling but not the scaling itself, confirming the prediction of Sec.~\ref{secProductAnsatz}. Since the dissipator only depends on $\{ \xi_{j} \bmod{2\pi} \}$, the conclusion in this study therefore holds for any shift of locations by integer multiples of wavelength $\lambda_{0}$.  

\section{Spontaneous Spin Ordering in a Regular Lattice}\label{SpinOrderingRegularLattice}

\begin{figure}[htp]
	\center
\includegraphics[width=\columnwidth]{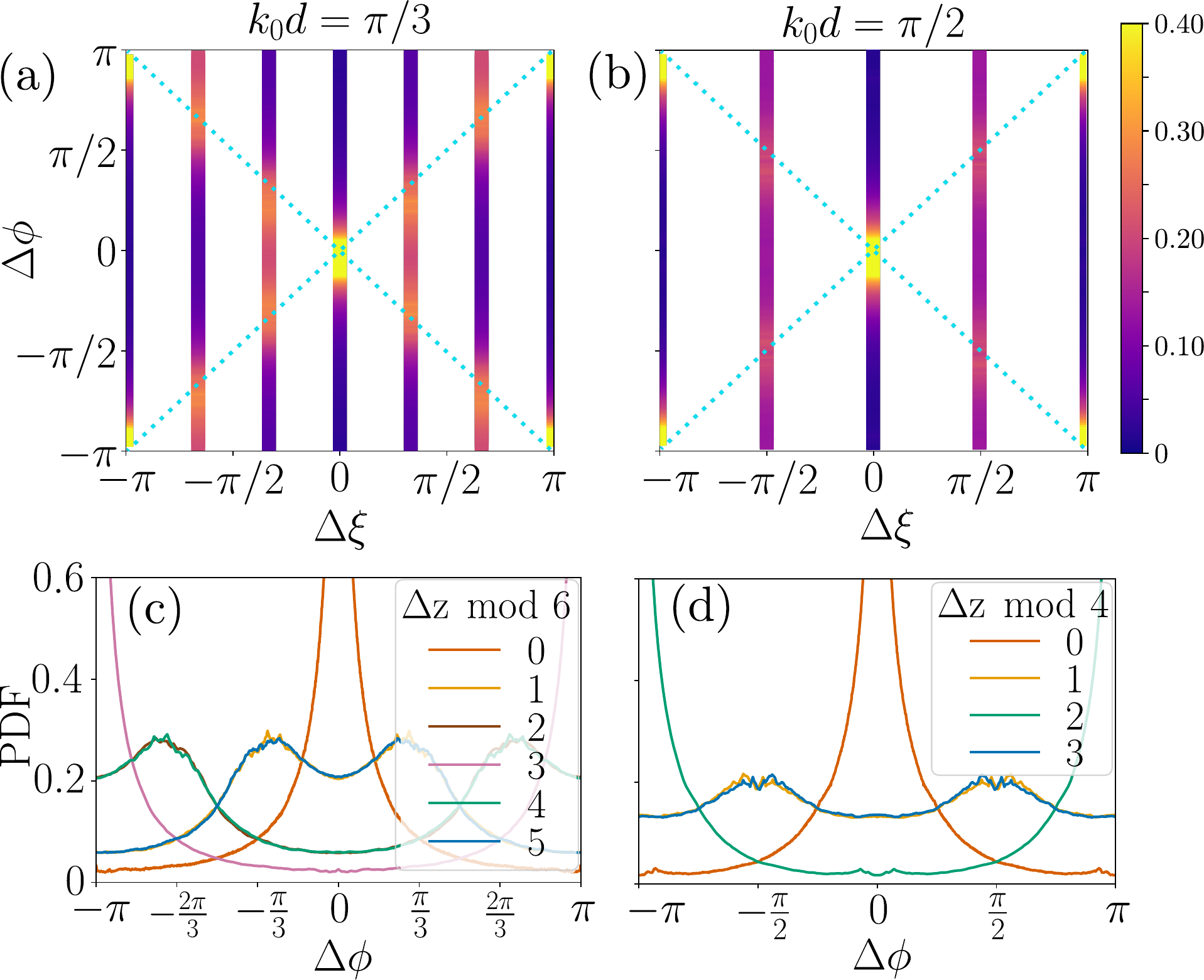}
	\caption{Spontaneous spin ordering in regular lattices. (a) and (b) Plot of the distribution $p(\Delta \xi, \Delta \phi)$ of the relative propagating phase $\Delta \xi$ and relative spin orientation $\Delta \phi$ for $k_0 d = \pi/3$ and $k_0 d = \pi/2$, respectively. Note that for a regular array, each slice in $\Delta z$ is infinitely narrow, but has been broadened for better visibility. The plots in (c) and (d) show the corresponding probability distribution function (PDF) of $\Delta \phi$ at a fixed distance $\Delta z$. The color bar for (a) and (b) shows the corresponding PDF in (c) and (d). See the Supplemental Material \cite{SupMat} for similar results for other lattice constants. Simulations use $N=400$ atoms and $10^3$ trajectories.}
	\label{FigRegularLattice}
\end{figure}

In addition to the disordered atomic arrays considered in Sec.~\ref{SubSecSpinOrdering}, in this appendix we show that a similar spontaneous spin ordering exists for regular lattices with a lattice constant $d$. In this case, the upper bound can be written as 
\begin{equation}\label{RBoundRegular}
\mathbb{R} = \max_{ \{ \phi_{j} \} } \sum_{i,j} \frac{\gamma}{4} \cos \left[ \left(i-j\right) k_{0}d \right]  \cos( \phi_{i} - \phi_{j} ).
\end{equation}
For $k_0 d \leq \pi/(2N)$, we know $\left|i-j\right| k_{0}d \leq \pi/2$, which makes $\{ \phi_{i} = \phi_{j} \}$ the optimal configuration. As discussed in Sec.\ref{SubSecSpinOrdering}, for a larger value of $k_0 d$, we expect the approximate bound $\mathbb{R}_{<}$ to provide an accurate estimate for the maximal decay rate. It predicts an ordering of the atomic dipoles according to  $\{ \phi_{j} = \phi_{1} +  (j-1) k_{0}d \}$ or $\{ \phi_{j} = \phi_{1} - (j-1) k_{0}d  \}$. Then a lower estimate of the upper bound is given by 
\begin{equation}
    \mathbb{R}_{<} = \sum_{i,j} \frac{\gamma}{4} \left(\cos \left[ \left(i-j\right) k_{0}d \right]\right)^2.  
\end{equation}
Specifically, we expect ordering patterns similar to spin waves, with a period of $2\pi/(k_{0}d)$. In Fig.\,\ref{FigRegularLattice}, we show the distribution of the relative angles $\Delta \phi$ at distance $\Delta z$ between a pair of spins, for different lattice constants $d$. There are peaks around $\Delta \phi \approx \pm \Delta z k_0 d$ as predicted from the product state ansatz. It can be seen that the ferromagnetic and antiferromagnetic orderings are much more pronounced than other configurations. Interestingly, distributions $\Delta \phi \approx \pm \pi/2$ are almost flat compared to other cases. This is because, in Eq.~\eqref{RBoundRegular}, the terms with $(i-j) k_{0} d \approx \pm \pi/2$ vanish, which makes the value of $\Delta \phi$ arbitrary in the optimal configuration.

\section{Time Evolution of Left- and Right-Emission Rates}\label{MoreData}

\begin{figure}[tb]
	\center
\includegraphics[width=\columnwidth]{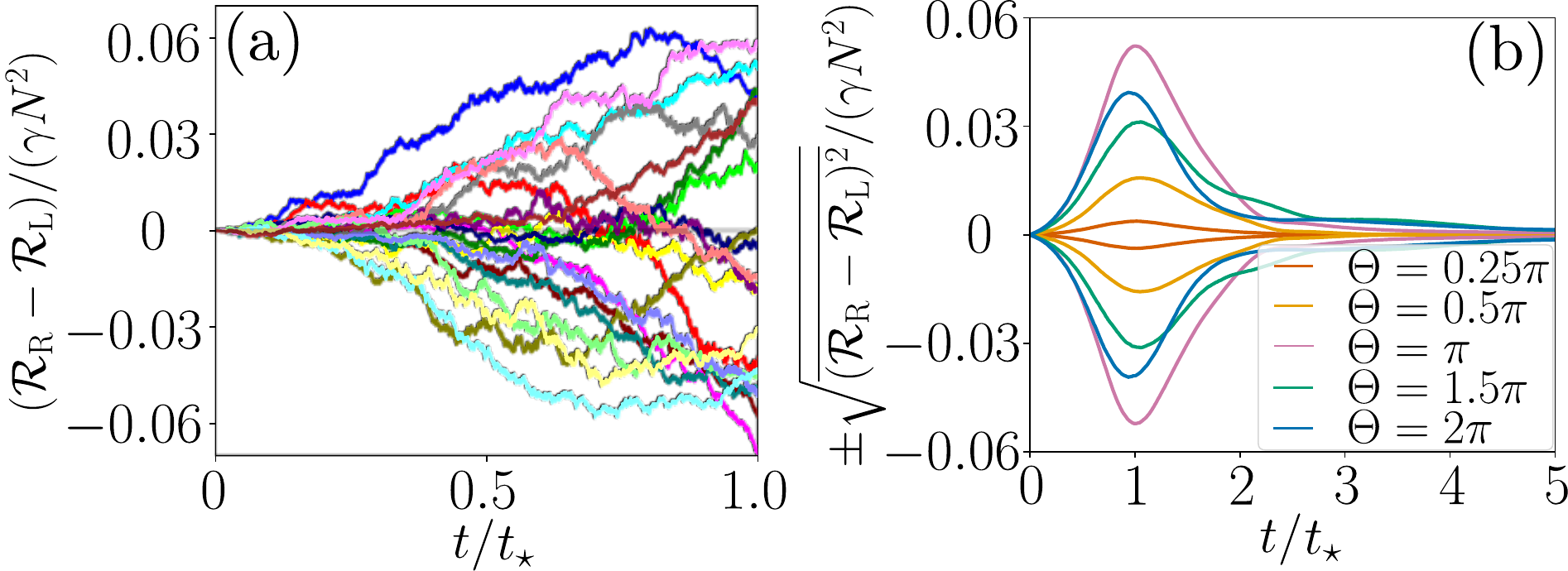}
	\caption{Evolution of left-right difference in each QSDMF trajectory. (a) $20$ example quantum trajectories for $\Theta = 2\pi$. (b) The standard deviation of $(\mathcal{R}_{\text{R}} - \mathcal{R}_{\text{L}})$ for different disorder strengths.}
	\label{FigEvolutionRRRL}
\end{figure}

As a complement to Fig.~\ref{FigRRRL},  Fig.~\ref{FigEvolutionRRRL}(a) shows the time evolution of the left-right emission asymmetry $(\mathcal{R}_{\text{R}} - \mathcal{R}_{\text{L}})$ for individual QSDMF trajectories for $\Theta = 2\pi$. It can be seen that most trajectories exhibit an asymmetry in the emission rate with a preference for either one or the other side, which is consistent with the fact that the auto-correlations are larger than the cross-correlations. Already in the early stage of dynamics, an emission into one direction, which imprints a phase pattern on the spins, is likely to seed consecutive emissions into the same direction and thus preserve this ordering pattern in the time evolution. However, this asymmetry is not completely fixed during the decay, and some of the trajectories change their preferred direction of emission during the decay process. As shown in Fig.~\ref{FigEvolutionRRRL}(b), the relative strength of the asymmetry peaks at $t_{\star}$ and increases with disorder strength.

\section{Boundary Effects}\label{BoundaryEffect}

\begin{figure}[tb]
	\center
\includegraphics[width=0.5\columnwidth]{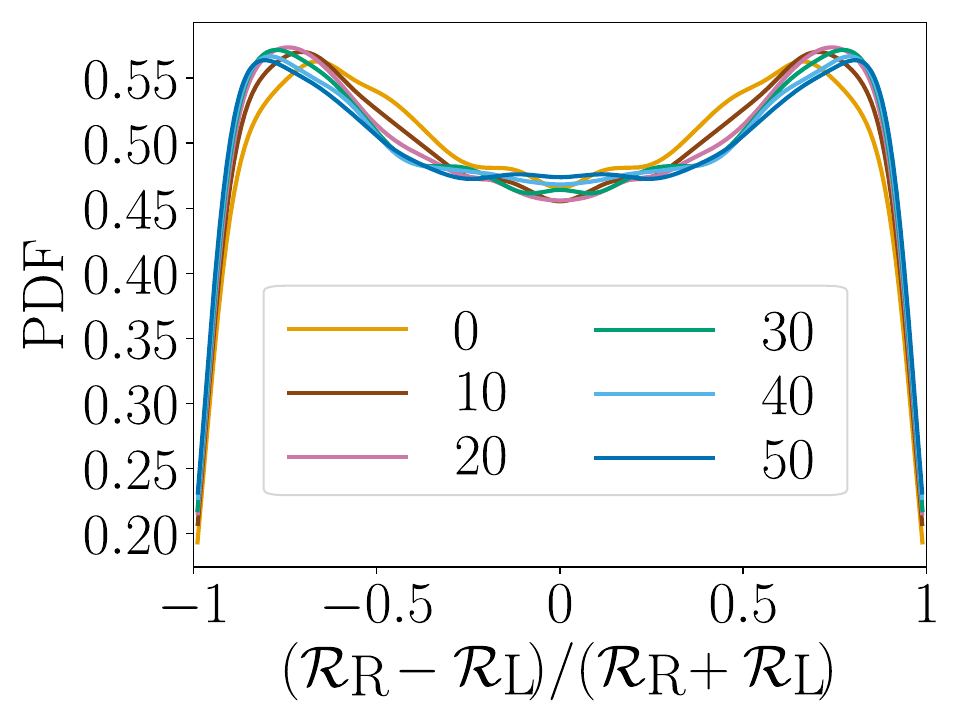}
	\caption{Probability distribution of the decay-rate asymmetry obtained from QSDMF trajectories, after removing the data from the last $n_{\text{B}}=0,10,20,30,40,50$ atoms at each boundary, i.e.\,keeping only the contribution from the bulk $(N-2 n_{B})$ atoms. As $n_{\text{B}}$ increases, only contributions from the bulk remain. The two peaks correspond to the two spontaneous spin-ordering configurations, while the boundary atoms introduce only minor modifications to the distribution. Simulations for this plot have been performed with $N=400$, $\Theta=2\pi$, and using $5\times10^{4}$ trajectories.
    }
	\label{FigPDFAsymmetry}
\end{figure}

In a recent work~\cite{CardenasLopez2025}, the authors investigate closely related spin-ordering effects in the steady states of actively pumped atomic arrays and observe a strong influence of the boundaries: spins at the left (right) boundary tend to favor the anticlockwise (clockwise) configurations. As a result, boundary atoms contribute approximately symmetrically to emission into the left- and right-propagating modes.

To quantify the influence of boundary atoms on spin ordering in the current setting, Fig.~\ref{FigPDFAsymmetry} shows the probability distribution of the decay-rate asymmetry $\mathcal{A} = (\mathcal{R}_{\text{R}} - \mathcal{R}_{\text{L}}) / (\mathcal{R}_{\text{R}} + \mathcal{R}_{\text{L}})$, which is computed after artificially removing the contributions of $n_{\text{B}}$ atoms from each boundary. The distribution exhibits two pronounced peaks, corresponding to the two bulk spin-ordering configurations $\rho^{\pm}$. As $n_{\text{B}}$ increases, the contribution of the boundary atoms is progressively suppressed, leading to a slight enhancement of the two peaks, but no significant changes. These results show that while boundary atoms do exhibit behavior distinct from that of the bulk, their effect on the decay-rate asymmetry is minor. The dominant contribution arises from the bulk atoms, which realize the two spontaneous spin-ordering configurations predicted in Sec.~\ref{SubSecSpinOrdering}.

\bibliography{Bib.bib, BibFootnotes}


\newpage

\clearpage

\appendix

\onecolumngrid

\clearpage

\setcounter{equation}{0}
\setcounter{figure}{0}
\setcounter{table}{0}
\setcounter{page}{1}

\makeatletter

\@removefromreset{equation}{section}
\@removefromreset{figure}{section}
\@removefromreset{table}{section}

\renewcommand{\theequation}{S\arabic{equation}}
\renewcommand{\thefigure}{S\arabic{figure}}
\renewcommand{\thetable}{S\arabic{table}}

\providecommand{\theHequation}{}
\providecommand{\theHfigure}{}
\providecommand{\theHtable}{}

\renewcommand{\theHequation}{supp.equation.\arabic{equation}}
\renewcommand{\theHfigure}{supp.figure.\arabic{figure}}
\renewcommand{\theHtable}{supp.table.\arabic{table}}

\renewcommand{\appendixname}{}

\makeatother


\begin{center}
	
	{\large\bf Supplemental Material for \\``Robust Superradiance and Spontaneous Spin Ordering in Disordered Waveguide QED''}
	
	\vspace{0.5cm}
	
	Xin  H.  H.  Zhang,$^{1,2,3}$ Daniel Malz$^{4}$ and Peter Rabl$^{1,2,3}$

    {\it 
$^{1}${Technical University of Munich, TUM School of Natural Sciences, Physics Department, 85748 Garching, Germany} \\
$^{2}${Walther-Mei{\ss}ner-Institut, Bayerische Akademie der Wissenschaften, 85748 Garching, Germany}\\
$^{3}${Munich Center for Quantum Science and Technology (MCQST), 80799 Munich, Germany} \\
$^{4}${Department of Mathematical Sciences, University of Copenhagen, Universitetsparken 5, 2100 Copenhagen, Denmark}
}
\end{center}





\section{Supplementary Plots}

\begin{figure}[h!]
	\center
\includegraphics[width=\columnwidth]{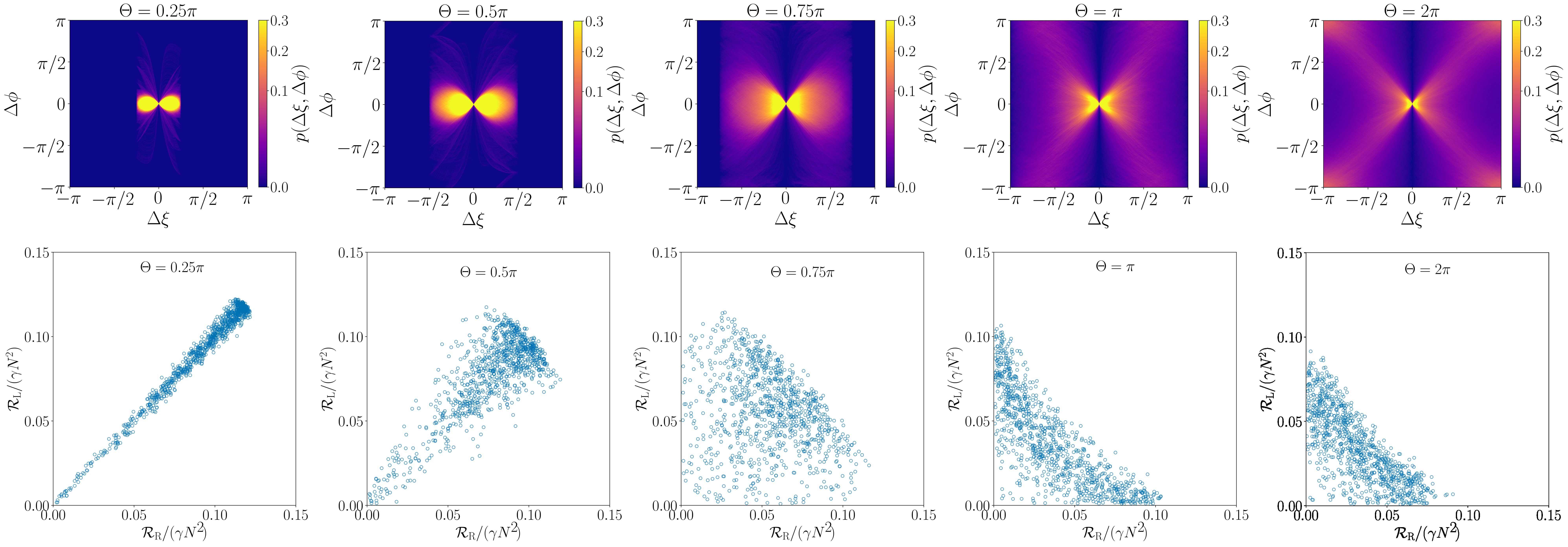}
	\caption{Changing of spin ordering as disorder strength $\Theta$ increases, in complement to Fig.~\ref{FigSpinAlignment} and \ref{FigRRRL}. Upper: Joint distribution $p(\Delta \xi, \Delta \phi)$ at superradiant time $t_{\star}$ for $10^3$ trajectories with $N=400$. Lower: Joint distribution of $(\mathcal{R}_{\text{R}} ,\mathcal{R}_{\text{L}})$  for $10^3$ trajectories with $N=400$. The clockwise and anticlockwise spin orderings gradually become more and more evident as $\Theta$ increases. Then, we can see that the fluctuation around the mirror-symmetric emission also increases gradually as the disorder strength increases.   }
	\label{FigMoreRRRL}
\end{figure}

\begin{figure}[h!]
	\center
\includegraphics[width=0.6\columnwidth]{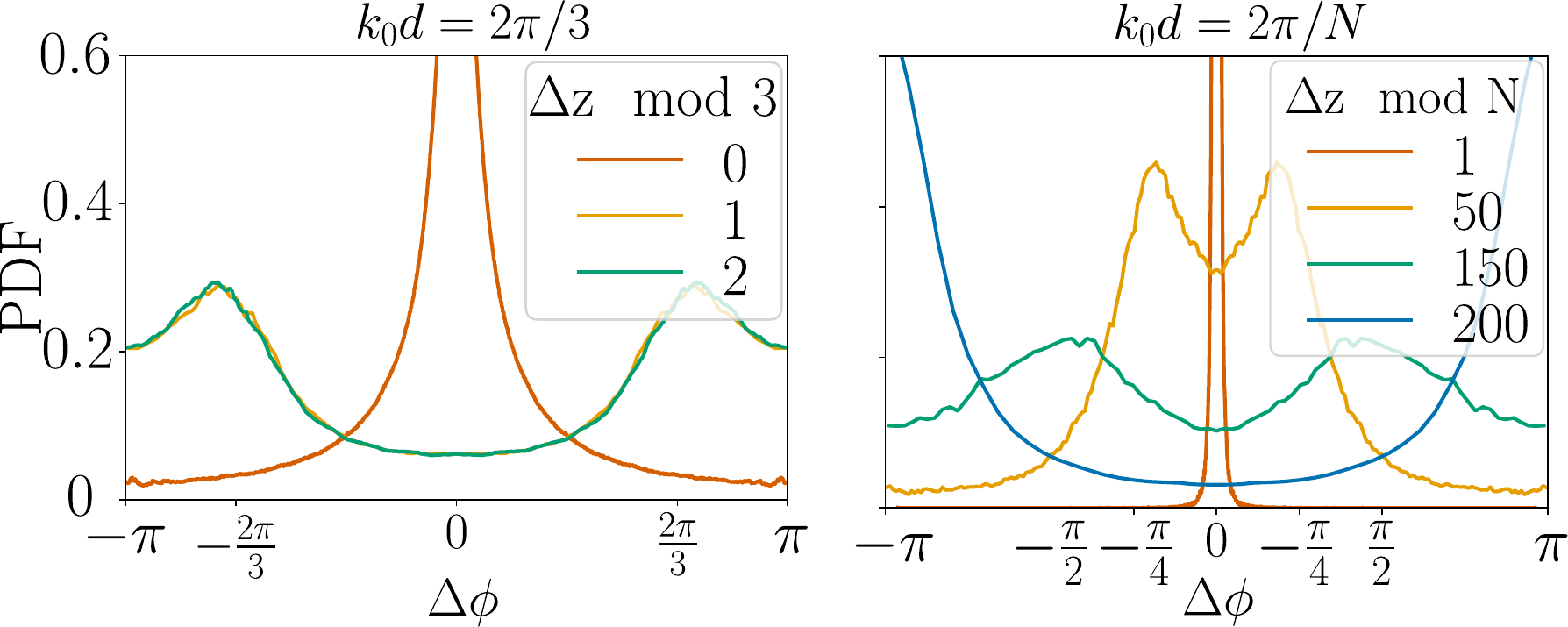}
	\caption{Spontaneous spin ordering in regular lattices, complementary to Fig.~\ref{FigRegularLattice}. Left and right panels show the probability distribution function of the relative spin orientation $\Delta \phi$ at a fixed distance $\Delta z$, for $k_0 d = 2\pi/3$ and $k_0 d = 2\pi/N$. Simulations use $N=400$ atoms and $10^3$ trajectories.}
	\label{FigMoreRegularLattice}
\end{figure}

\begin{figure}[h!]
	\center
\includegraphics[width=0.6\columnwidth]{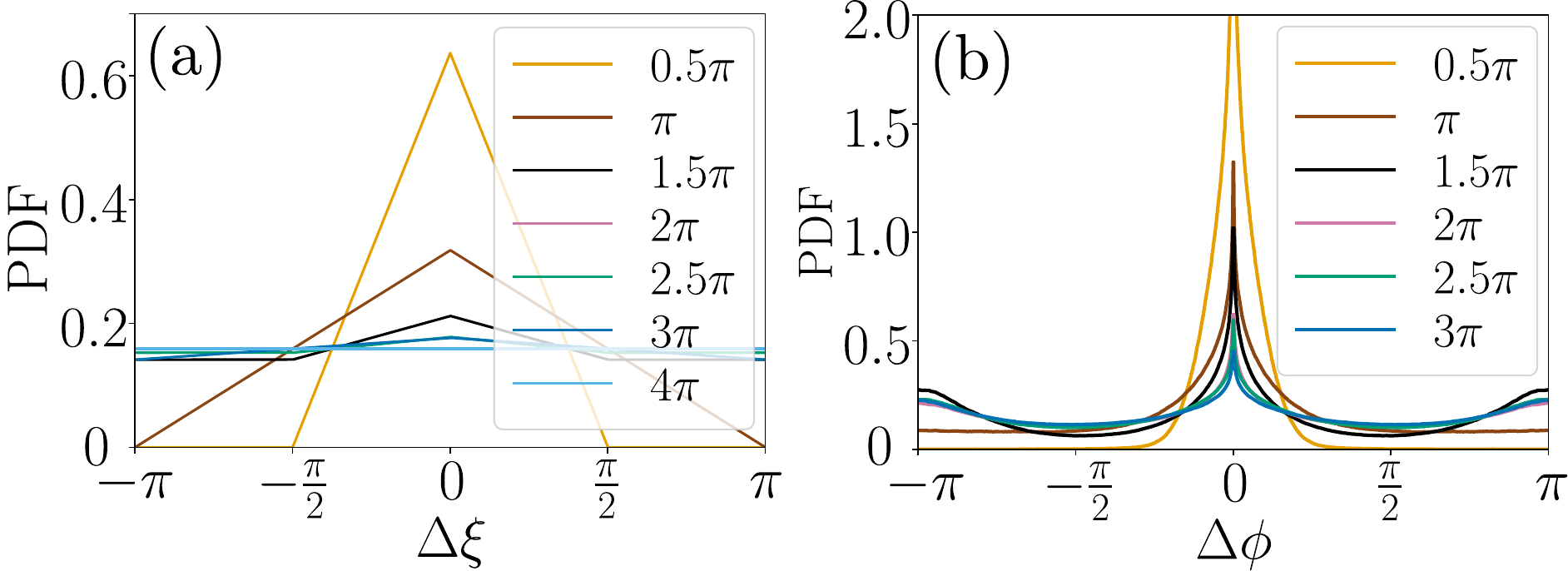}
	\caption{(a) Probability distribution of relative propagating phases $\Delta \xi = (\xi_i - \xi_j) \bmod 2\pi$, which is a wrapped triangular distribution. (b) Probability distribution of relative spin angles $\Delta \phi = (\phi_i - \phi_j) \bmod 2\pi$, for $N=400$ and $10^3$ trajectories. When $\Theta = 0$, the system is permutation-invariant in QSDMF trajectories, and all spins align along the same direction, resulting in $p(\Delta \phi) = \delta(0)$ as expected. As $\Theta$ increases, the relative propagating phase $\Delta \xi$ becomes more uniformly distributed in $(-\pi,\pi)$. Accordingly, $p(\Delta \phi)$ also becomes more evenly distributed, meaning the global spin alignment of the atoms weakens. Instead, a new spin ordering emerges, with spins aligning according to their position-dependent phases as revealed by the joint probability distribution $p(\Delta\xi, \Delta\phi)$ shown in Sec.~\ref{SubSecSpinOrdering}. Additionally, we can observe that, under strong disorder, there is a weak tendency for alignment and anti-alignment in the distribution $p(\Delta\phi)$. This can be attributed to the fact that the joint distribution $p(\Delta\xi, \Delta\phi)$, shown in Fig.~\ref{FigSpinAlignment}, has less spreading around $\Delta \xi \approx 0$ and $\pi$. For similar reasons as in Appendix \ref{SpinOrderingRegularLattice}, the ordering around $\Delta \phi \approx \Delta \xi \approx \pm \pi/2$ is weaker than ordering in other directions.}
	\label{FigMoreXiPhiandRDiff}
\end{figure}

\begin{figure}[h!]
	\center
\includegraphics[width=0.5\columnwidth]{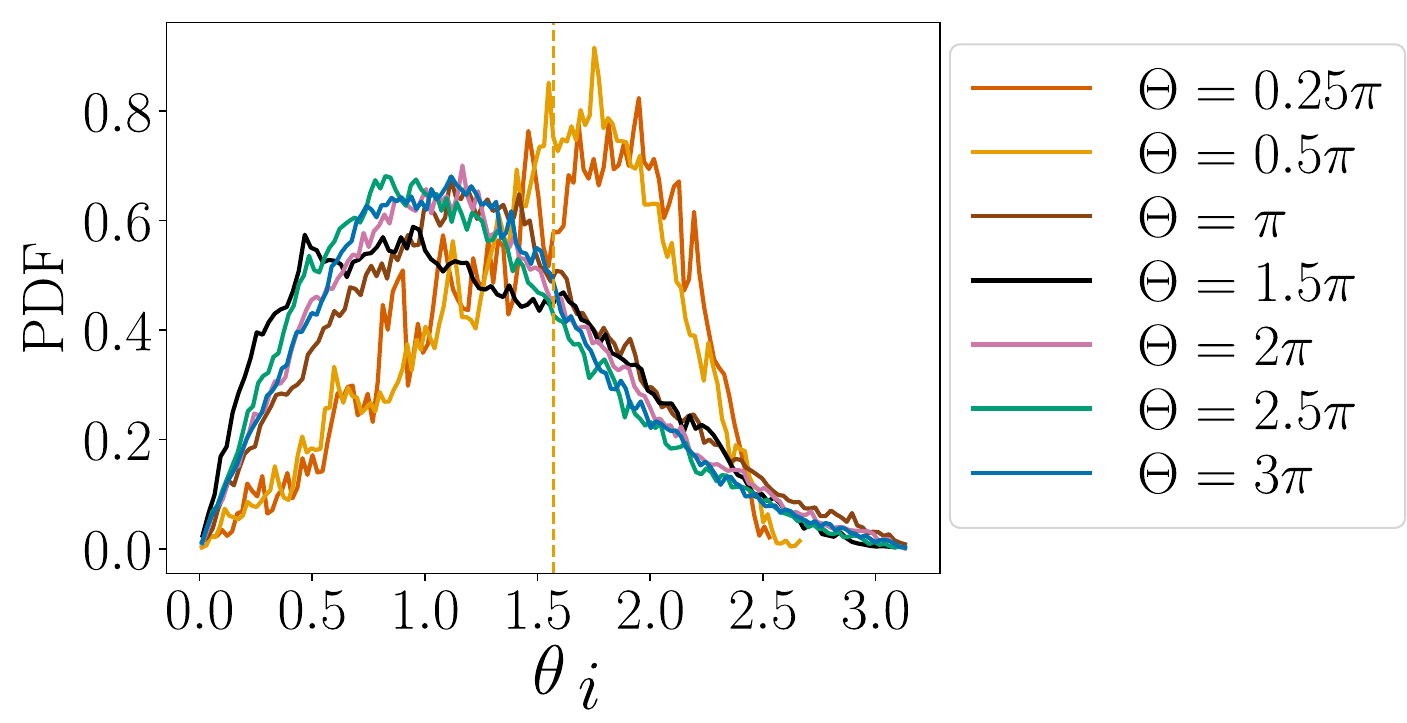}
	\caption{Probability distribution of each spin angle $\theta_i$ at $t_{\star}$ for different disorder strengths $\Theta$. The upper bound in Sec.\,V.\,A. predicts a distribution around $\theta = \pi/2$. We see that $\theta_i$ is indeed distributed around $\pi/2$ for small disorder $\Theta \leq \pi/2$. However, the center of the distribution shifts to the left for larger $\Theta$. Both the finite-width distribution of $\theta_i$ and the finite-width distribution in the $\phi$ direction in Fig.~\ref{FigSpinAlignment} contribute to the discrepancy between the upper bound $\mathbb{R}_{<}$ and the numerical data. Simulations use $N=400$ atoms and $10^3$ QSDMF trajectories. }
	\label{FigPDFTheta}
\end{figure}

\begin{figure}[h!]
	\center
\includegraphics[width=0.9\columnwidth]{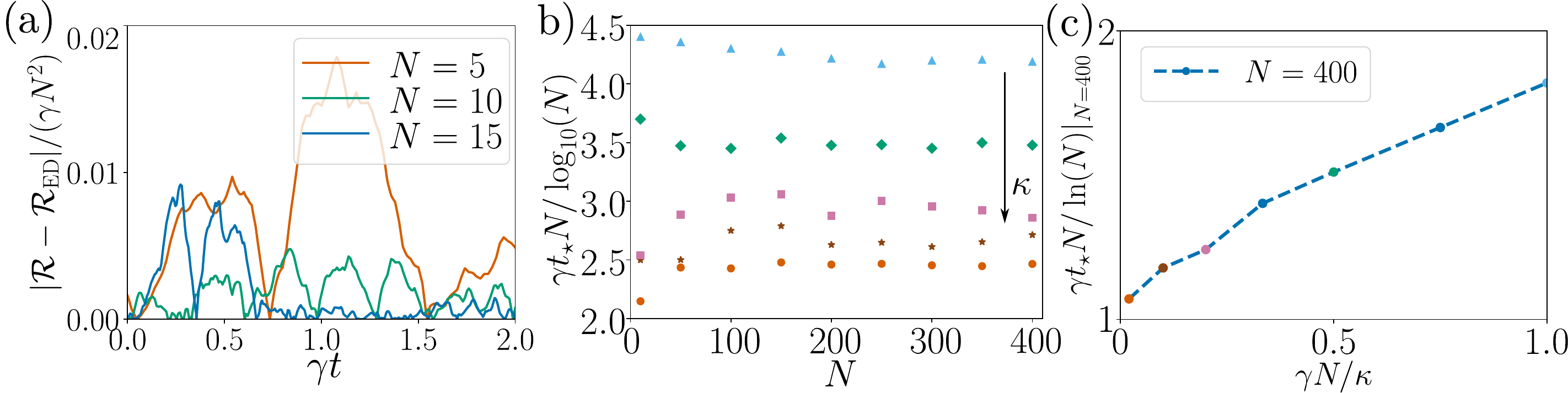}
	\caption{In complement to Fig.~\ref{FigNMDecay}, (a) Error of DTWA for $N=5,10,15$ atoms coupled with one lossy cavity, (b) Burst time $t_{\star}$ vs.\,$N$ for atoms coupled with two cavities, for $\kappa/(\gamma N) = 1, 2, 5, 10, 50$ (from top to bottom), and (c) The roughly linear relation between $\gamma t_{\star} N/\ln(N)$ and $\gamma N/\kappa$ for $N=400$. $t_{\star} \sim \ln(N)/(\gamma N)$ scaling persists in the weakly non-Markovian regime, with the prefactor decreasing to $1$ as $\kappa$ increases. Simulations use $10^3$ trajectories.}
	\label{FigMoreNM}
\end{figure}

\begin{itemize}
    \item Fig.~\ref{FigMoreRRRL} shows the smooth crossover between weak ($\Theta=\pi/4$) and strong disorders ($\Theta=2\pi$). 
    \item Fig.~\ref{FigMoreRegularLattice} shows plots of spin ordering for a regular lattice with $k_{0} d= 2\pi/3$ and $2\pi/N$.
    \item Fig.~\ref{FigMoreXiPhiandRDiff} shows probability distributions of the propagating phases $\Delta \xi$ and the relative spin angles $\Delta \phi$. 
    \item Fig.~\ref{FigPDFTheta} shows probability distributions of the spin angle $\theta$ at superradiant times, for different disorder strengths $\Theta$. 
    \item Fig.~\ref{FigMoreNM} shows the error of DTWA in the non-Markovian case and the scaling of $t_{\star}$ there.
\end{itemize}

\newpage


\section{Reduction of the SDE for Efficient Treatment of Cavity Dynamics}

By setting $\alpha_{\text{R}/\text{L}} = \alpha_{\text{R}/\text{L}}^{(0)} + \alpha_{\text{R}/\text{L}}^{(1)} $, the stochastic differential equation (SDE) in Eq.~\eqref{AlphaSDE} can be separated into a stochastic part
\begin{equation}\label{alpha0EOM}
\begin{split}
    d \alpha^{(0)}_{\text{R}/\text{L}} &= - \frac{\kappa}{2} \alpha^{(0)}_{\text{R}/\text{L}} dt + \sqrt{\frac{\kappa}{2}} dW_{\text{R}/\text{L}},
\end{split}
\end{equation}
and a deterministic part
\begin{equation}\label{alpha1EOM}
\begin{split}
    \frac{d}{dt} \alpha^{(1)}_{\text{R}/\text{L}} &= - \frac{\kappa}{2} \alpha^{(1)}_{\text{R}/\text{L}} + \frac{g}{\sqrt{2}}  \tilde{J}_{\text{R}/\text{L}}. \\
\end{split}
\end{equation}
For the initial state, we can let $\alpha^{(0)}_{\text{R}/\text{L}}(0) = \alpha_{\text{R}/\text{L}}(0)$ and $\alpha_{\text{R}/\text{L}}^{(1)}(0) = 0$. In numerical simulations, the stochastic part \eqref{alpha0EOM} can be solved using the Euler-Maruyama algorithm with a very small time step, to generate $\mathcal{N}$ stochastic trajectories of $\alpha^{(0)}_{\text{R}/\text{L}}$ to high precision. Then, for each trajectory of $\alpha^{(0)}_{\text{R}/\text{L}}$, we can input it as dynamical variables into the set of deterministic EOMs of spin variables in Eq.~\eqref{SpinEOM} and variables $\alpha^{(1)}_{\text{R}/\text{L}}$ in \eqref{alpha1EOM}. The resulting deterministic ordinary differential equations (ODE) can then be solved to high precision with higher-order integrators. This can make the higher-order integrators of ODE directly applicable to the dynamics of the spin variables.

\section{Irrelevance of Detuning in DTWA}\label{AppendixIrrelevanceDetuningDTWA}

In Sec.\,\ref{secDisorderedAtomFrequency}, we show the numerical results that detuning is irrelevant as $N$ gets sufficiently large in DTWA simulations. The argument there uses the assumption that the cumulant expansion of finite order is accurate. Here we show that, since each trajectory in DTWA is treated at the MF level, no such additional assumption is needed in this analysis. In the DTWA formalism, after the unitary transformation for the rotating frame, the EOMs \eqref{SpinEOMAfterElimination} in Appendix \ref{DTWAappendix} become
\begin{eqnarray}
\label{SpinEOMAfterEliminationRotating}
    d \tilde{s}_{j}^{-} ={}& -& \frac{\gamma}{2} \tilde{s}_{j}^{-} dt + \frac{\gamma}{2} \sum_{l} s_{j}^{z} \tilde{s}_{l}^{-} e^{i|\xi_{j} - \xi_{l}|} e^{i(\delta\omega_{j} - \delta\omega_{l}) t} dt \nonumber\\
    &+& \sqrt{\frac{\gamma}{2}} s_{j}^{z} dW_{j} e^{i\delta\omega_{j}t}   \\
        d s_{j}^{z} ={}& -& \gamma s_{j}^{z} dt - \gamma \sum_{l} \left( \tilde{s}_{j}^{+} \tilde{s}_{l}^{-} e^{i|\xi_{j} - \xi_{l}|} e^{i(\delta\omega_{j} - \delta\omega_{l}) t} + \text{c.c.} \right) dt  \nonumber\\
        & -& \sqrt{2\gamma} \left( \tilde{s}_{j}^{+} dW_{j} e^{i\delta\omega_{j}t} + \text{c.c.} \right) \nonumber,
\end{eqnarray}
with $\tilde{s}_{j}^{\pm} (t) = s_{j}^{\pm}(t) e^{\mp i \delta\omega_{j} t}$. As expected, this is just the Heisenberg EOM \eqref{SpinEOMHeisenberg} plus noises. Similar to Sec.\,\ref{secDisorderedAtomFrequency}, the detuning gives corrections $\sim \Delta_\omega t \sim  (\Delta_\omega/\gamma) \ln(N)/N$ to the EOMs, which become negligible for $N/\ln(N) \gg \Delta_\omega/\gamma$. Since DTWA is accurate for this problem, this also indicates that the required cumulant order $C \ll N/\ln(N)$ for superradiance. The analysis using DTWA and that using CE mutually corroborate the conclusion that finite detuning is irrelevant for large $N$ in superradiance.

\section{Comment on Experimental Relevance}\label{AppendixExpRelevance}

Here, using the parameters of a few recent experiments, we discuss possible experimental realizations and tests of our theoretical predictions. Some experiments already exhibit strong spatial disorder. Experiments with weak spatial disorder can be used to study our predictions for regular arrays.  

\begin{itemize}
    \item In Refs.~\cite{LiedlPRX2024,BachArXiv2024}, $\sim 1000$ cesium atoms are chirally coupled to a 1D nanophotonic waveguide. Since the atoms are probabilistically loaded into the trapping potential and the trapping lattice spacing is incommensurate with the resonant wavelength, the resulting atomic separations give random propagation phases that are roughly uniformly distributed in $(0,2\pi)$ \cite{MengPRL2020}. Therefore, this setup corresponds to the strongly disordered case with $\Theta \approx 2 \pi$ in our study. Although random propagation phases can be absorbed into a redefinition of the Pauli operators for a chiral waveguide and thus do not affect the dynamics, coupling to a bidirectional waveguide can be realized with nonchiral waveguide or different atomic transitions and used to directly probe our theoretical predictions.

\item In the microwave regime, Ref.~\cite{BrehmNatMat2021} realizes 8 superconducting qubits coupled to a 1D waveguide. The qubits are densely spaced, with separations much smaller than the resonant wavelength and a spatial disorder strength $\Theta \ll 1$. Therefore, spatial disorder can be neglected in this setup, while frequency disorder may play a more important role. When scaled to more qubits, this setup can be used directly to probe our predictions with a predefined disorder pattern or ordering behavior of regular lattices discussed in Appendix \ref{SpinOrderingRegularLattice}.

\item Ref.~\cite{TiranovScience2023} studies 3 quantum dots coupled to a photonic crystal waveguide. The positional uncertainty in this setup is $\sim \lambda/10$, i.e.~$\Theta \sim \pi/5$, which lies in the weakly disordered regime for this small number of emitters.

 \item As an outlook for future study, we mention collective emission into 3D free space. Recently, the experiment in Ref.~\cite{FerioliNatPhys2023} studied a 3D pencil-shaped cloud of $\sim 2000$ laser-cooled $^{87}$Rb atoms under external driving. The size of the atomic cloud is $\sim 0.5\lambda$ and $\sim 20\lambda$ along the two axes. This gives large disorder strengths $\Theta \sim \pi$ and $40\pi$ in these two directions. Our theory can potentially be extended to this quasi-1D case and probed in such experiments.

\end{itemize}


\end{document}